# Interaction between Surface Waves on Wire Lines

D.Molnar[1]*, T. Schaich[1], A. Al Rawi[1],[2] and M.C. Payne[1]

[1]Theory of Condensed Matter Group, Cavendish Laboratory, University of Cambridge, CB3 0HE, UK

[2]BT Labs, Adastral Park, Orion building, Martlesham Heath, IP5 3RE, UK

*corresponding author, dm611@cam.ac.uk

**Abstract**

**This paper investigates the coupling properties between surface waves on parallel wires. Finite Element Method (FEM) based and analytical models are developed for both single wire Sommerfeld and Goubau lines. Starting with the Sommerfeld type wave, we derive the analytical expression based on the assumption that the two-wire surface wave is a superposition of the two surface waves on the individual wires. Models are validated via measurements and a comparison study conducted between the analytic and the FEM-based computations for coupled Sommerfeld type lines. We then investigate the coupling between two Goubau lines with the FEM model. The measurement and calculations show remarkable agreement. The FEM-based and the analytical models match remarkably well too. The results exhibit new properties of favourable effects on surface waves propagation over multiple conductors. The short range behaviour of the coupled wires and consequently the existence of an optimum separation of coupled wires, is one of the most significant findings of this paper. We comment on the relevance of our results especially in relation to applications of high bandwidth demands and immanent cross-coupling effects.**

Introduction

Electromagnetic surface waves (SW) of various kinds are recently attracting considerable attention due to potential for many applications from telecommunication to plasmonics ( [1] [2] [3] [4]). Importantly, these emerging technologies have also been recently studied as backhaul solutions to the network standard 5G [5], [6] . In this paper, we focus on SW on single wire conductors and particularly on the coupling effects between parallel single wire conductors.

Cylindrical SWs, as a solution to Maxwell's equations were first proposed by Sommerfeld over a century ago ( [7], [8]), and these solutions were reexamined by Goubau and others from the 1950s [9] [10].They found that it is possible to ensure a propagating SW mode as long as the wire has finite conductivity (Sommerfeld case), or if corrugations are introduced on the surface of the wire or if the wire is coated by a dielectric. The two latter modifications lower the effective conductivity of the conductor, making the mode well confined to the surface [9]. The introduction of corrugations is known in the modern literature as 'spoof plasmons' [11] and applying a dielectric coating has become known as the Goubau or G line.

Standalone or single wire waveguides have been investigated since their inception and their propagation characteristics have been studied in detail ( [9] [12] [13] [14] [15] ). Recently, interest has been rekindled due to the possible application of these single conductor lines in large bandwidth applications, owing to their near dispersion-less and low loss characteristics [16]. Moreover, twisted copper wire pairs carrying SWs have also been proposed as possible alternatives to fibre networks for high-speed, multi Gigabit/sec data transmission rates [17] . The interest in reusing the telephone lines is motivated by the substantial cost benefits compared to replacing them by fibre-optics. Without loss of generality in our studies we use the dimensions and material parameters specific for the UK telephone network [18]. However, our qualitative conclusions should equally apply to wires which have different dimensions from the ones considered in this work.

In this work, we investigate the surface wave coupling effects and report our results for two parallel wires, both for the Sommerfeld and the Goubau wires. We first employ the small wire assumption (weak coupling) and then compare the results with finite element method (FEM) numerical calculations (strong coupling). In the latter no explicit assumptions are made about the wires' separation or their diameters. We will concentrate on the frequency range of up to 300GHz, which is relevant for the latest technologies and, most notably, will be in use for the 5G network [19].

**Results**

Standalone (Single) Wire Waves: First we focus on the single wire waves and we use the analytically calculated solutions to validate our FEM results. We study wires with dimensions and material properties adopted from [18] throughout this work as follows: wire radius (a) =0.5mm, wire radius including coating (b) =0.55mm, conductivity ($\sigma$) =5.57e7 S/m, relative permittivity ($\varepsilon_d$) =2.54 and loss tangent (tan$\delta$) =1e-4.

A schematic illustration of the wire and the cylindrical coordinate system, as well as the 2D cross section and the cartesian coordinate systems (x,y,z) used in the FEM study for both type of wires are shown in Figure 1.

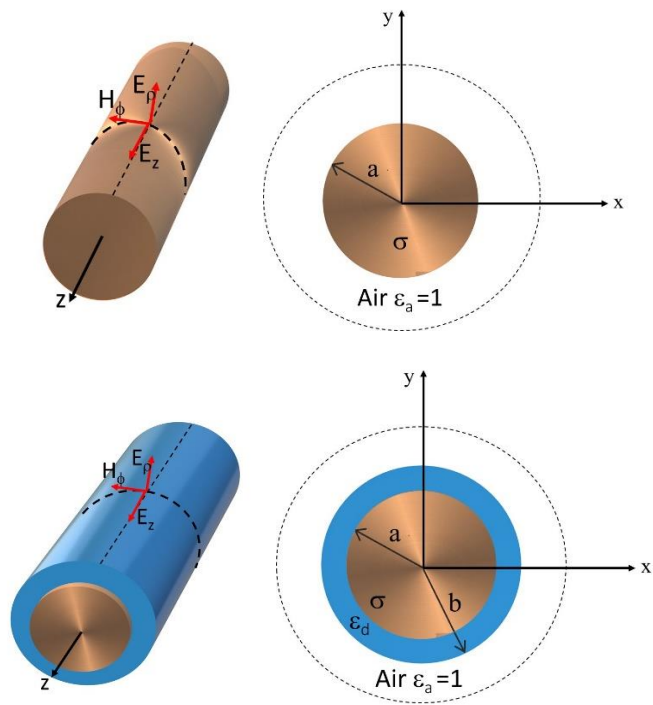

*Figure 1 Schematic illustration of the Sommerfeld wire (uncoated) and the Goubau (coated) wire, showing the coordinate system used in this paper. The cross sections are also shown. The wire radius* a *and the wire radius plus coating thickness* b *are shown in the cross section. The material properties, the conductivity of copper, $\sigma$, the relative permittivity of air, $\varepsilon_a$ and relative permittivity of the dielectric coating of the wire, $\varepsilon_d$ are also shown.*

**Sommerfeld wire wave**

The matching of the electric and magnetic field components, $E_z$ and $H_\phi$ on the surface of the conductor, (i.e. at ρ=a ), yields the two conditions from which the characteristic equation can be found [7]-[9]. We use an iteration scheme based on Orfanides [20] to find the complex roots of equation (1) with Matlab . $J_0$, $Y_0$ are the $0^{th}$ order Bessel functions of the first and second kind, and $H_0^{(1,2)}$ are the Hankel functions $0^{th}$ order of $1^{st}$ and $2^{nd}$ kind [21] [22], $H_0^{(1,2)}(\gamma\rho) = J_0(\gamma\rho) \pm iY_0(\gamma\rho)$, where the + and – signs denote the $1^{st}$ and $2^{nd}$ order Hankel function respectively.

$$\frac{\gamma}{\varepsilon_a}\frac{H_0^{(1)}(\gamma a)}{H_1^{(1)}(\gamma a)} = \frac{\gamma_c}{\varepsilon_c}\frac{J_0(\gamma_c a)}{J_1(\gamma_c a)} \qquad (1)$$

Further the lateral wave vector in air γ and in the conductor $\gamma_c$ are $\gamma = \sqrt{k_0^2\varepsilon_a - \beta^2}$, $\gamma_c = \sqrt{k_0^2\varepsilon_c - \beta^2}$, where β is the longitudinal wave vector and $k_0$ is the free space wavevector, and $\varepsilon_a$ and $\varepsilon_c$ are the relative permittivities of the air and conductor, respectively.

**Goubau wire wave**

In the case of the Goubau wire, the equations can be similarly derived from the continuity of the corresponding field components and therefore a characteristic equation can be found as shown in equation (2) [9] [12]. We use a similar iteration scheme as for the Sommerfeld wire wave to obtain the solution without any losses (dielectric or metallic conductive) with Matlab.

$$\frac{h}{\varepsilon_d}\frac{Z_0(hb)}{Z_1(hb)} = -\frac{\gamma}{\varepsilon_a}\frac{K_0(\gamma b)}{K_1(\gamma b)} \qquad (2)$$

Where the function Z is expressed as $Z_i(h\rho) = J_i(h\rho) - \frac{J_0(ha)}{Y_0(ha)}Y_i(h\rho), i = 1,2$.

Further the lateral wave vector in air γ and in the dielectric h are $\gamma = \sqrt{k_0^2\varepsilon_a - \beta^2}$

and $h = \sqrt{k_0^2\varepsilon_d - \beta^2}$. Where $\varepsilon_a$ and $\varepsilon_d$ are the relative permittivities of the air and the dielectric coating, respectively.

Once the lossless solution is found for either the Goubau or Sommerfeld type wave it can be used to obtain the longitudinal attenuation along the line (α) perturbatively by equation (3):

$$\alpha = \frac{P_{loss}}{2P_{TR}} \quad (3)$$

where $P_{loss}$ is the power loss per unit wire length, and $P_{TR}$, is the transmitted power by the SW.

**Comparison between FEM and analytically calculated results for single wire waves**

We have solved the Helmholtz equation using the commercial software COMSOL Multiphysics, to obtain the propagation characteristics of the surface wave (see Supplementary Information for more details). We compare our results obtained with the FEM solver to the results obtained from the semi-analytic (or from now on for brevity, analytic approach) for the standalone wire case. We focus on the dominant $E_{00}$ mode only. We calculate with COMSOL the complex propagation constant (β), and from it the attenuation $\alpha$, i.e. the imaginary part of $\beta$, and the wavenumber ($\eta$), i.e. the real part of $\beta$, as a function of frequency in the range of 50GHz-250GHz. These quantities can be compared with the analytically obtained results. The graphs are plotted for both the Sommerfeld wire wave (uncoated wire) and the Goubau wire wave (coated wire). Excellent agreement is found between the FEM numerical models' results and the analytically obtained results. The relative difference is smaller than 1% in each case.

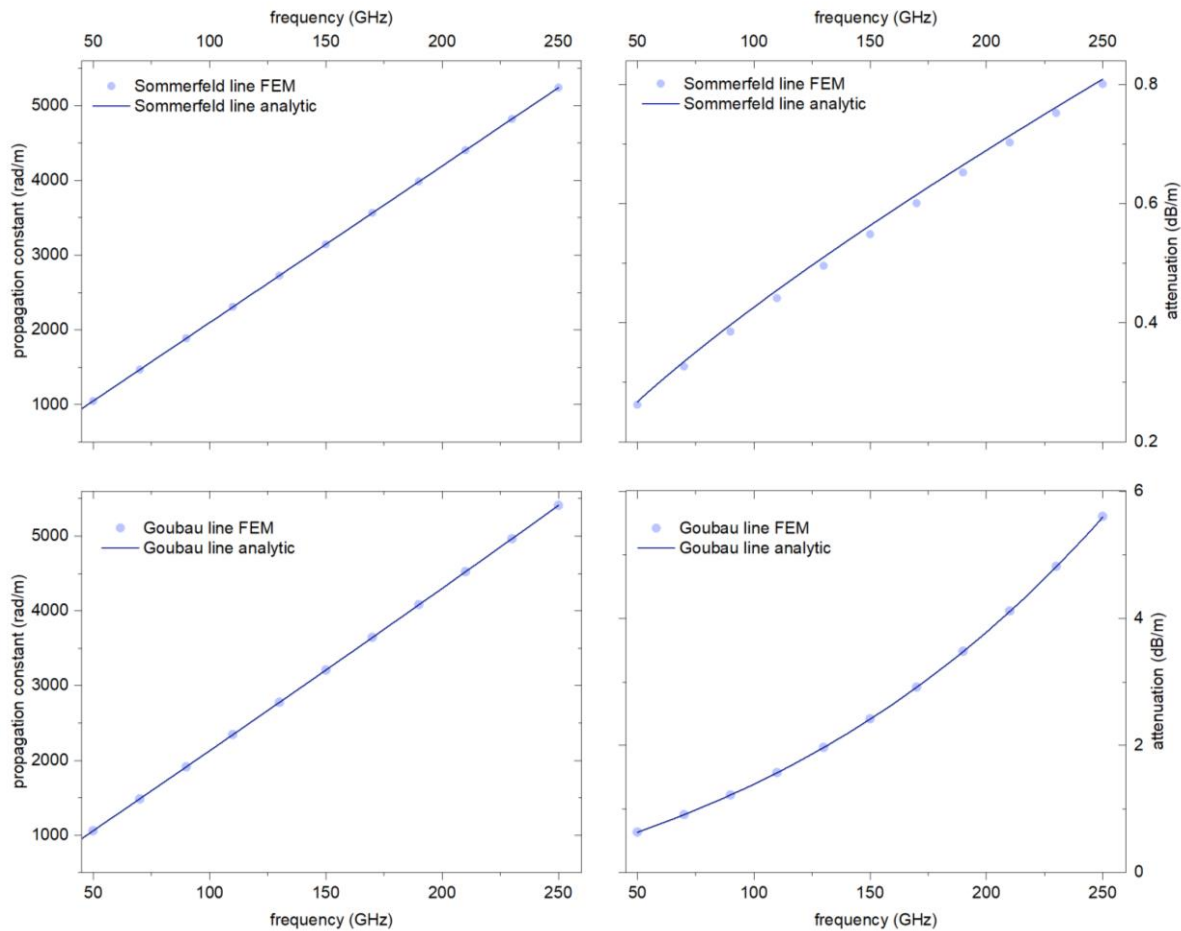

*Figure 2 Comparison of the FEM computed (light blue dots) and analytically calculated (solid blue line) propagation constants and attenuation as a function of frequency. Good agreement is found between the results obtained by the two different methods.*

Additionally, we also compare the field profiles, particularly the out of plane electric field ($E_z$) and the magnetic field ($H_\phi$) as a function of radius for both types of wire waves. In both cases the amplitudes are normalized to the value of $E_z$ and $H_\phi$ at the copper wire surface. The relative difference between the FEM numerical calculation obtained results and the analytic results is within less than 1% in both cases. These plots are shown in Figure 3 and Figure 4.

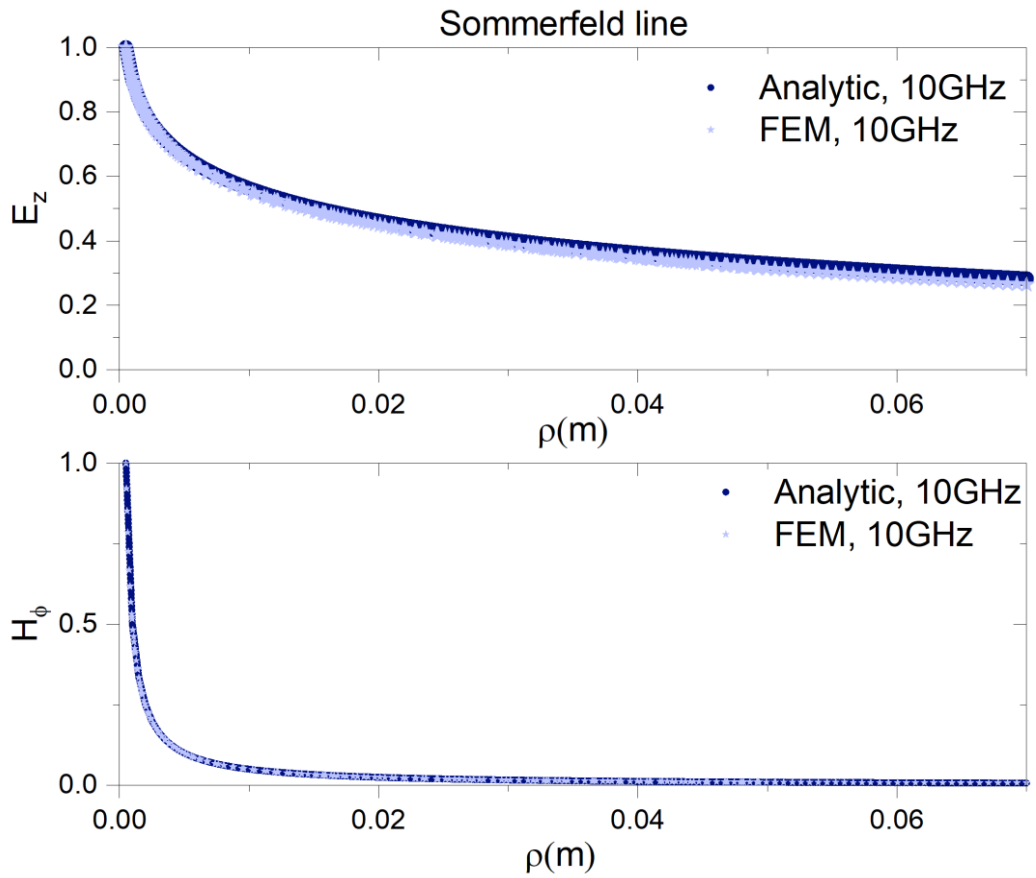

*Figure 3 Comparison of the FEM computed (light blue stars) and analytically calculated (solid blue) filed components $E_z$ (upper) and $H_\phi$ (lower) of the fields at 10GHz for the Sommerfeld line. The field components are plotted against the radial coordinate r and are normalized to their value at the conductor surface.*

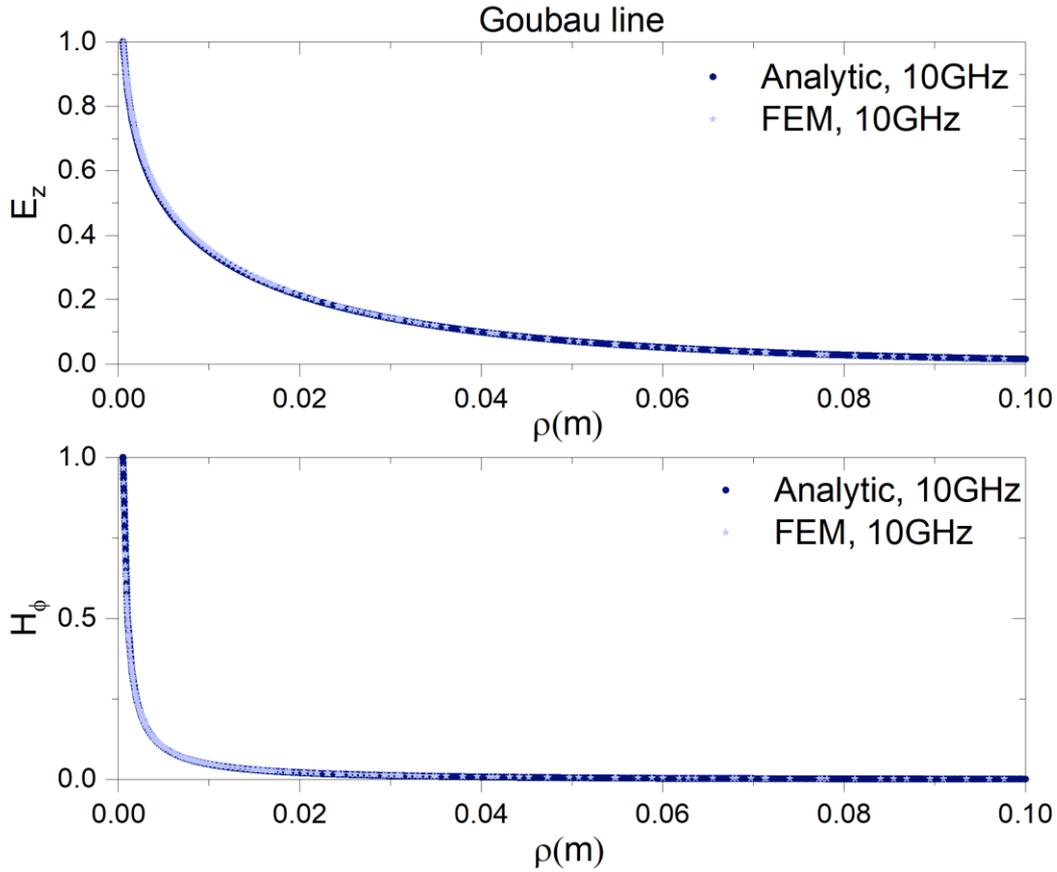

*Figure 4 Comparison of the FEM computed (light blue stars) and analytically calculated (solid blue) filed components $E_z$ (upper) and $H_\phi$ (lower) of the fields at 10GHz for the Goubau line. The field components are plotted against the radial coordinate r and are normalized to their value at the conductor surface.*

**Coupled surface waves**

The analysis of the standalone lines can be extended to two parallel lines carrying a surface wave. Meyerhoff first investigated the coupling of two weakly coupled SW lines [23], and then later Cook and Chu carried out theoretical work [24]. Their main interest was to extend Meyerhoff's approach and particularly to consider higher order modes and hybrid modes. In the paper they only reported the lowest order mode $TM_{00}$ coupling for two Goubau lines. More recently, Xu et al carried out numerical calculations for two coupled surface wave lines and focused on the characteristic impedance of the lines as well as reporting propagation constants. They found that the even modes have lower effective refractive index than the odd modes [25]. Furthermore, Xu et al only considered a fixed distance between the wires, and, therefore, fixed coupling strength. All of these works only considered the Goubau (coated wire) line. Little has been published on Somerfeld type lines' coupling characteristics. Additionally, Meyerhoff focused on the coupling between a single wire carrying a SW and a passive wire in the vicinity of it. Meyerhoff didn't investigate the scenario in which both wires carry a SW, the two wire wave case. In this work we investigate some of the propagation characteristics of the two wire wave and map their dependence on the relative distance between the wires. We focus on the lowest order symmetric (even) and antisymmetric (odd) modes of the dominant $E_{00}$ ($TM_{00}$) mode on each wire, and therefore the resultant two-wire wave.

We investigate the coupling between two wires that have their z axes parallel and are at distance d apart from one another. The geometry of the configuration is shown in Figure 5. Note that the wires are identical and have identical dimensions and material properties to the preceding sections.

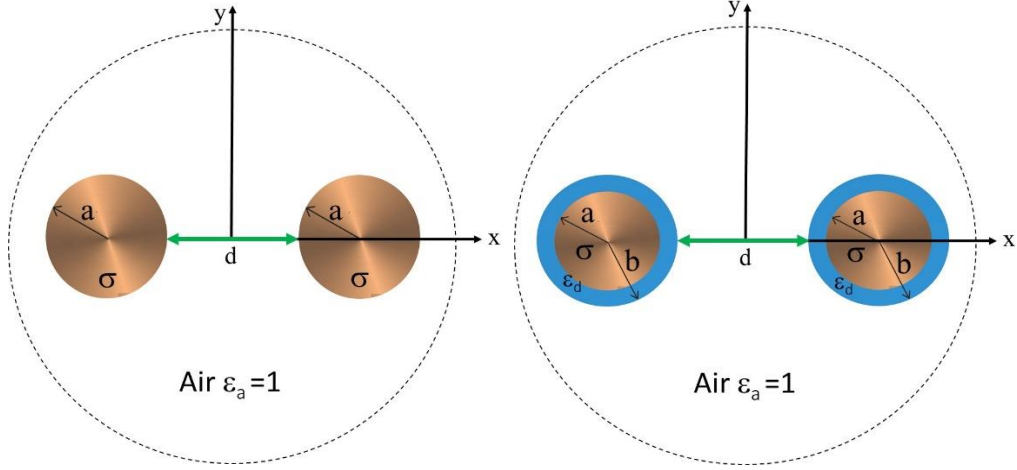

*Figure 5 Schematic illustration of the coupled Sommerfeld and Goubau wire waves. The separation distance between the wire waveguides,* d *is also shown. We vary* d *to study the propagation characteristics at each frequency as a function of d.*

**Weak Coupling -Analytic approach**

In this section, we derive the characteristic equation to find the complex propagation constant of the coupled wire wave. Note that for obvious reasons we will refer to this approach as analytic but in reality in the last step we performed of a numerical computation of the coupled characteristic equation, similar to the case of the single wire (see equations (1) and (2)). Here we base this assumption on Sommerfeld who first suggested this type of coupling but did not present any results [8]. The main assumptions are that the currents in the wires are azimuthally independent and the separation large between them, leading to the weak coupling limit. For a detailed derivation of this equation see Supplementary information.

The characteristic equation for the two wire wave modes is found to be:

$$\frac{\gamma}{\varepsilon} \frac{H_0(\gamma a) \pm H_0(\gamma(d+2a))}{H_1(\gamma a)} = \frac{\gamma_c}{\varepsilon_c} \frac{J_0(\gamma_c a)}{J_1(\gamma_c a)} \qquad (4)$$

Where + is for symmetric and – for antisymmetric mode two wire mode. This equation can be solved numerically by using Matlab and therefore the complex propagation constant $\beta$ of the coupled SWs can be obtained. For this analytic approach, we restrict our studies to that of the two Sommerfeld type waves, however the described method can be used to study the propagation characteristics of two parallel Goubau lines as well. The procedure is somewhat more involved in that case, as the first one assumes perfect metals and lossless dielectrics, and then in the second iteration the effects of these losses are taken into account. However, the assumptions of this analytic approach for the coupling in the limit of small separation distances between the wires is not realistic. Therefore, we use this method to compare it with our FEM results in the Sommerfeld type coupled waves and we will solve the Goubau coupled waves with FEM only.

**FEM Numerical Results**

Following the analytic approach for the coupling presented in the previous section we investigate the same configuration numerically, with the use of FEM. The boundary conditions are identical to the ones used in the standalone wire wave cases (see Supplementary Information), however now we utilize the symmetry inherent in the geometry for the two wire wave case [25], for further details see the Supplementary Information.

**Comparison between numerical and analytical results for coupled Sommerfeld wire waves**

We compare our results obtained by the analytic approach with the assumptions described previously with FEM computations. We expect the two results to asymptotically approach each other in the large separation limit. We did not stipulate any special assumptions on the separation of the wires or about the tangential field in the FEM models. Therefore, we are particularly interested in the differences that may be absent in the analytical results because of the assumptions taken.

For large separations, it can be seen in Figure 6 and Figure 7 that the results obtained by the methods are in very close agreement . However, for small separations the symmetric mode significantly differs from the results obtained analytically and those from FEM. In fact, the FEM calculation produces a characteristic shape of the attenuation curve with a clear, global minimum as a function of separation of the wires. This global minimum is located at a specific separation between the wires, which we will call the optimum distance. We expect this optimum distance to be a function of the frequency of the wave because the fields' radial extension is a function of frequency. We present the frequency dependence of this optimum distance later in the paper. The antisymmetric mode does not posses such characteristics. In the antisymmetric case the difference between the methods manifests itself in a larger predicted attenuation for small separations in FEM that for the analytic approach, but the overall shape is very similar obtained by either method and the numerical values are very close.

It can be appreciated immediately from the figures that the symmetric coupling case has much lower attenuation than the antisymmetric one. The difference is about 3dB/m at the smallest seperation distance and at the optimum separation distance for the symmetric mode is around 0.5dB/m. At very small separations in both cases we can observe a diverging attenuation, as the distance between the wires goes to zero. In addition, the antisymmetric mode's attenuation approaches a certain value from above, whereas the symmetric mode's attenuation approaches a certain value from below asymptotically for large separation distances. It is also clear that, apart from short separations both the symmetric and antisymmetric mode's attenuation asymptotically approach the values calculated by the analytic, weak coupling limit. As a matter of fact the deviation from it at small distances is the reason why we refer to the FEM calculated approach as the strong coupling limit. We leave the detailed explanation of these characteristics for the discussion section of this paper.

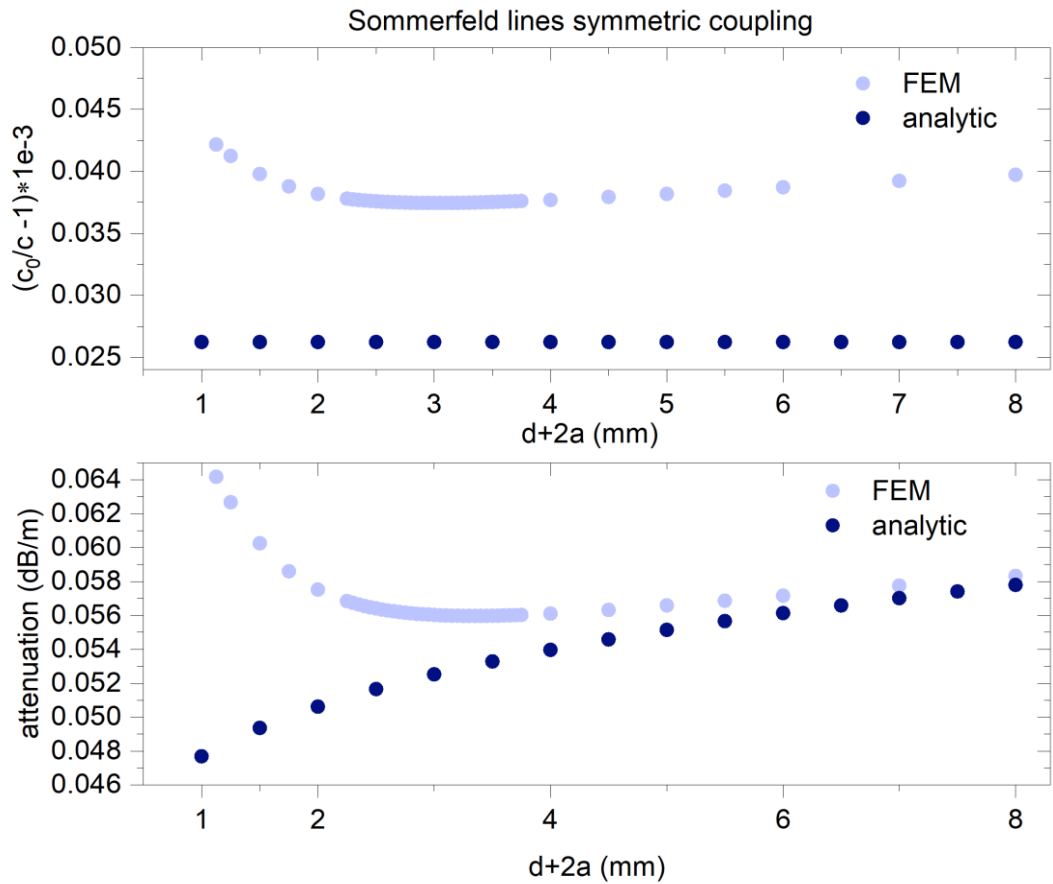

*Figure 6 Comparison between the analytic (dark blue) and FEM (light blue) obtained results for the symmetric mode's phase velocity and attenuation for two coupled Sommerfeld lines at 10GHz.*

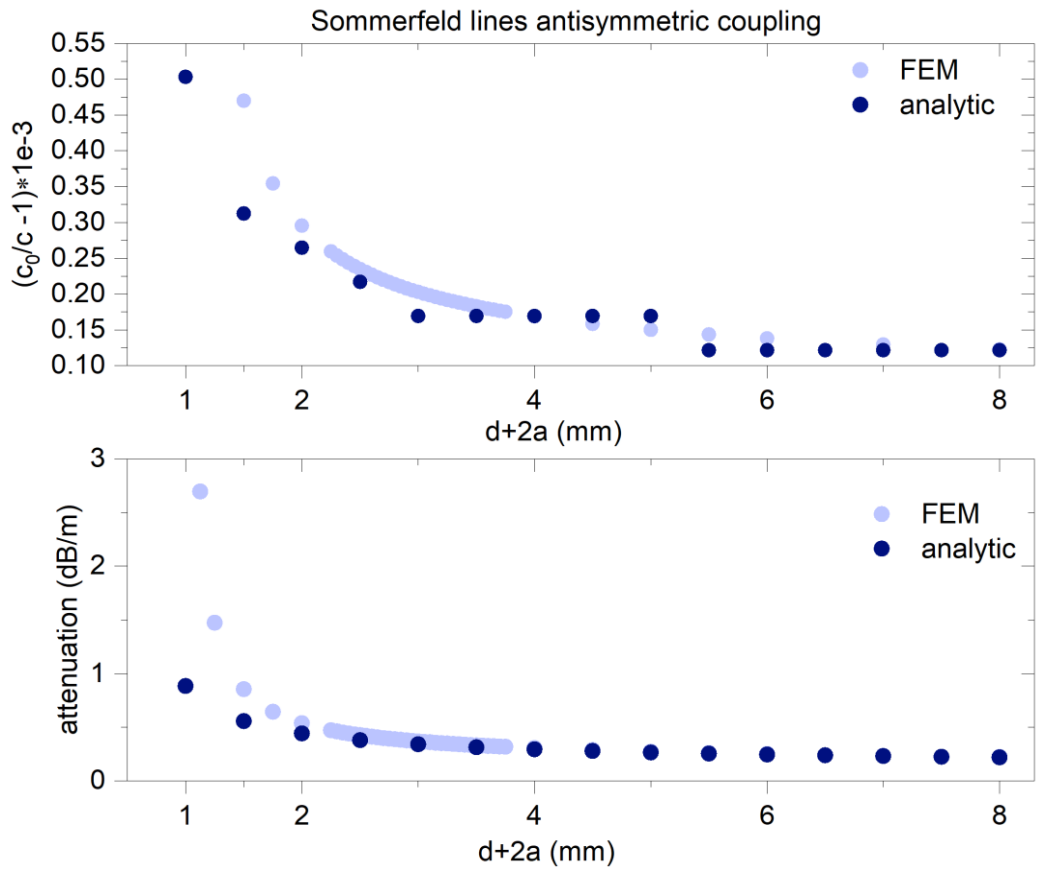

*Figure 7 Comparison between the analytic (dark blue) and FEM (light blue) obtained results for the antisymmetric mode's relative phase velocity and attenuation for two coupled Sommerfeld lines at 10GHz.*

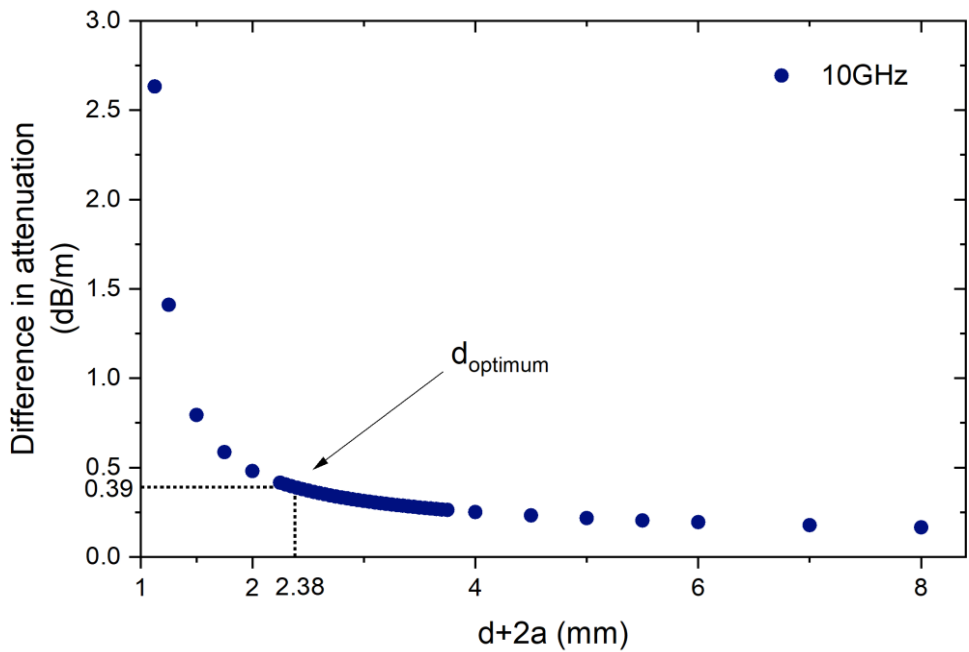

*Figure 8 The difference in attenuation between the symmetric and antisymmetric modes in dB/m at 10GHz for coupled Sommerfeld waves as a function of distance between the wire cores. The optimum distance for the symmetric mode at this frequency is also indicated on the graph*

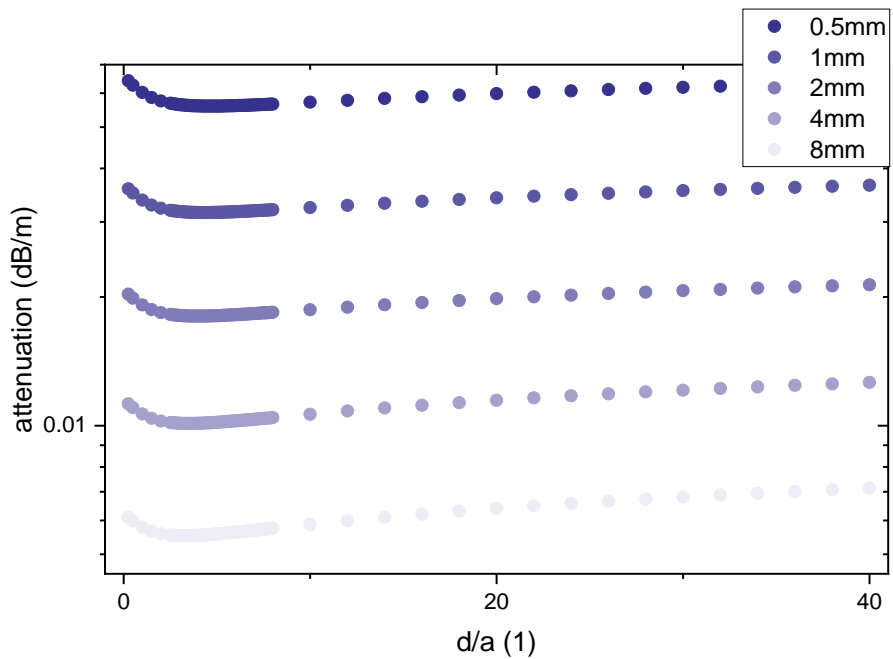

*Figure 9 Effect of radius on the attenuation for the symmetric mode in the case of two Sommerfeld wires. The attenuation is plotted in dB/m in log scale against the dimensionless factor of distance between the wires/radius of wires. These set of curves are produced at the frequency of 10GHz. Note that the attenuation minimum for each wire radius is located at the same d/a value. Increasing the wire radius to larger values results in a constant shift in the attenuation of the two wire wave as expected. This can be understood by considering the larger radius case as having effectively more*

*than one smaller radius two-wire wave as the available copper surface to bind the wave is increased. The optimum separation for the attenuation of the symmetric mode of the two-wire wave is a function of wire radius as outlined above. However it's constant in terms of the dimensionless parameter n, which is defined as the separation distance (d) divided by the wire radius (a).*

**Two Coupled Goubau Wire Waves**

We present the results of our FEM calculations of the symmetric and antisymmetric modes for the case of two axially parallel Goubau wires, each carrying a SW. The resulting attenuation and propagation constants at 10GHz are plotted in Figure 10 and Figure 11. The overall shape of the curves are very similar to that of the two coupled Sommerfeld wires. Importantly, the global minimum of the attenuation constant as a function of frequency still exists, therefore the optimum distance between the wires can be calculated for each frequency. For the Goubau wire wave the fields' radial extension at any given frequency is smaller than that of the Sommerfeld wire( with the same material and geometry parameters), hence the optimum distance is closer to 0mm at 10GHz than in the case of the uncoated wire.

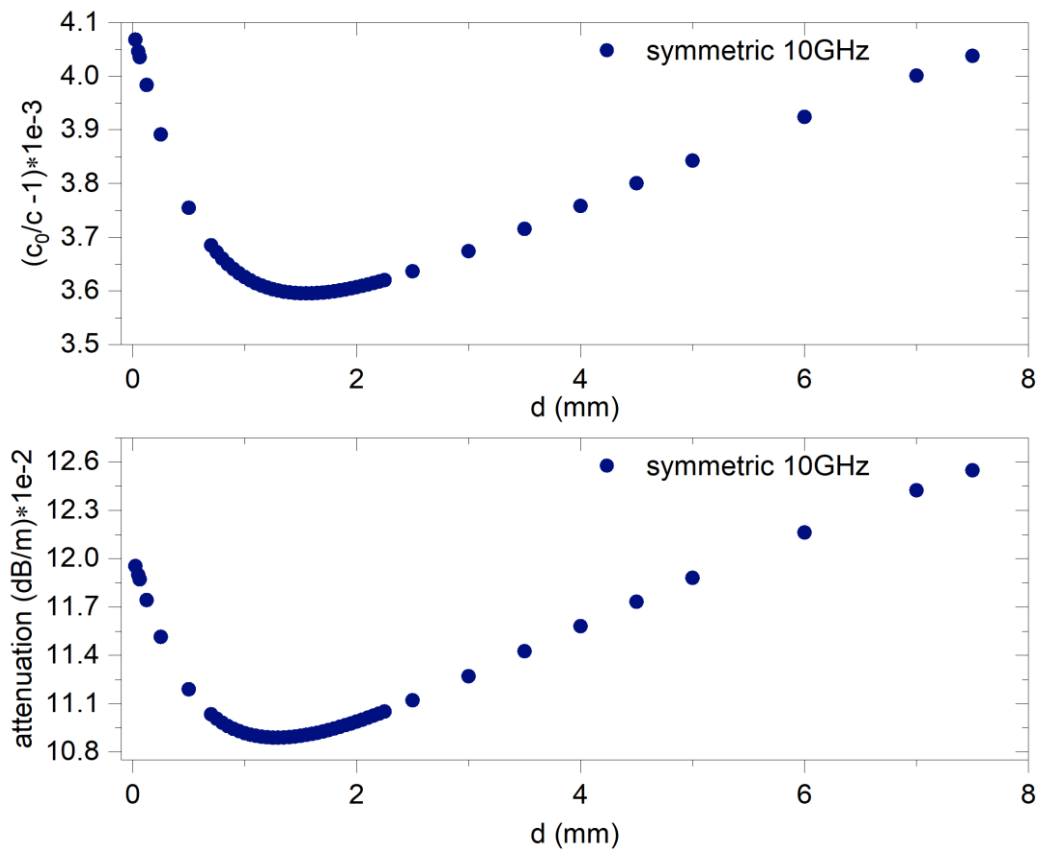

*Figure 10 Goubau two-wire wave relative phase velocity (upper panel) and attenuation (lower panel) for the symmetric mode at 10GHz as a function of separation between the two wires (d) in mm.*

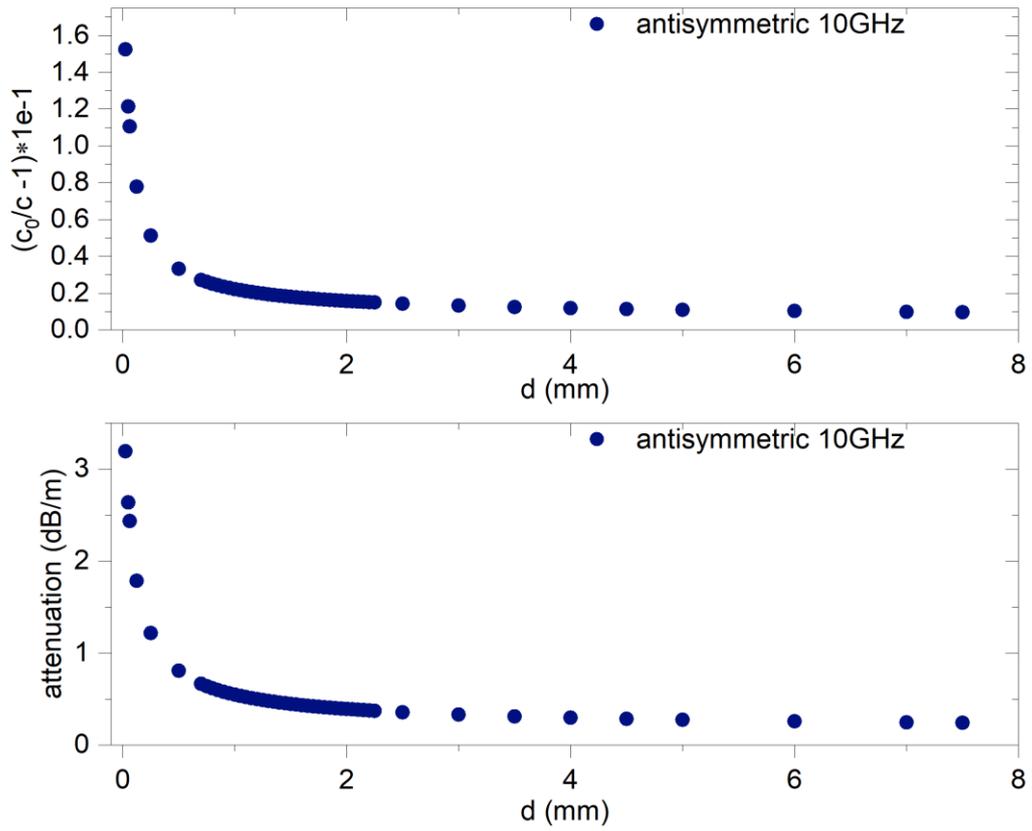

*Figure 11 Goubau two-wire wave relative phase velocity (upper panel) and attenuation (lower panel) for the antisymmetric mode at 10GHz as a function of separation between the two wires (d) in mm.*

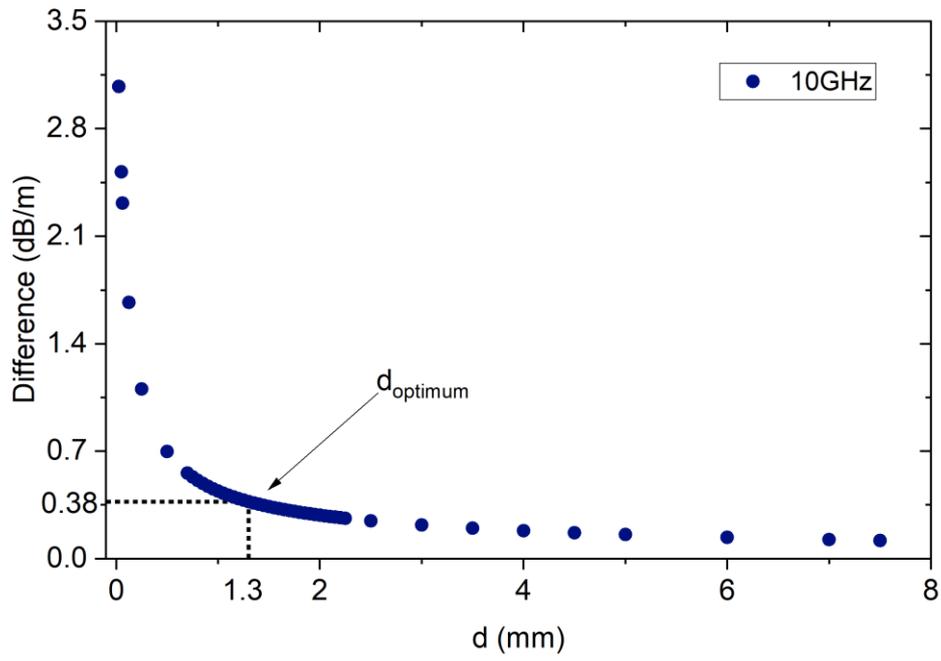

*Figure 12 Difference of attenuation between the differential and symmetric modes of the Goubau two wire wave at 10GHz. The optimum distance for the symmetric mode at this frequency is 1.3mm and the difference in attenuation between the modes is 0.38 dB/m.*

**Frequency dependence of the optimum distance between the wires**

Importantly, it can be concluded from the symmetric mode attenuation curve's shape that it is possible to find an optimum seperation distance of the wires from the perspective of longitudinal propagation loss for both two coupled Sommerfeld and Goubau wire waves. It is expected that this distance would decrease as a function of frequency as the radial extension of the fields is smaller for higher frequencies for both type of waves [20].

The fact that such an optimum separation exists for the symmetric mode is probably one of the most important findings of this paper. Therefore, we characterize the frequency dependence of this optimum distance in the symmetric mode up to 300GHz for both two coupled Sommerfeld and Goubau wire waves.

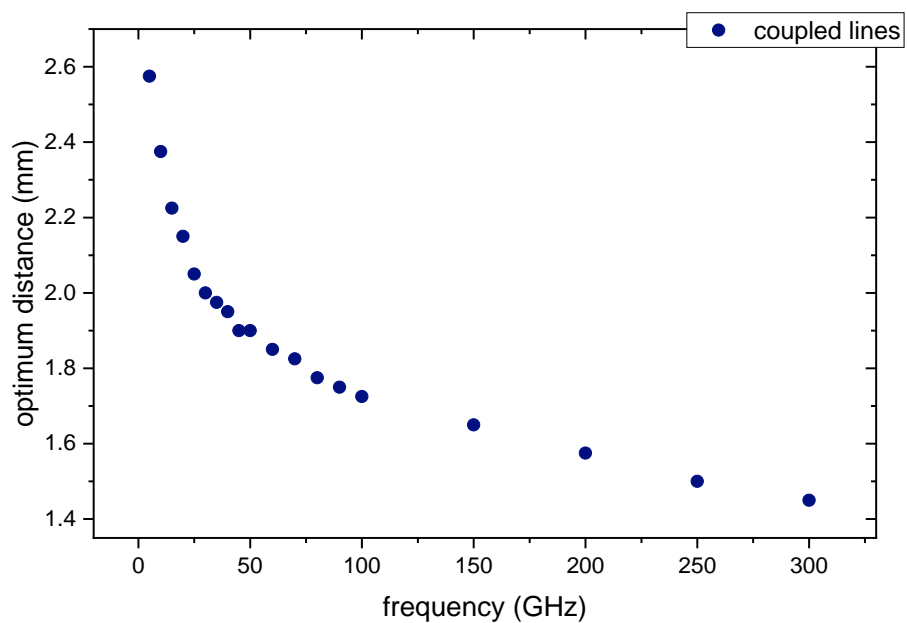

*Figure 13 Two wire Sommerfeld line's optimum distance in mm as function of frequency*

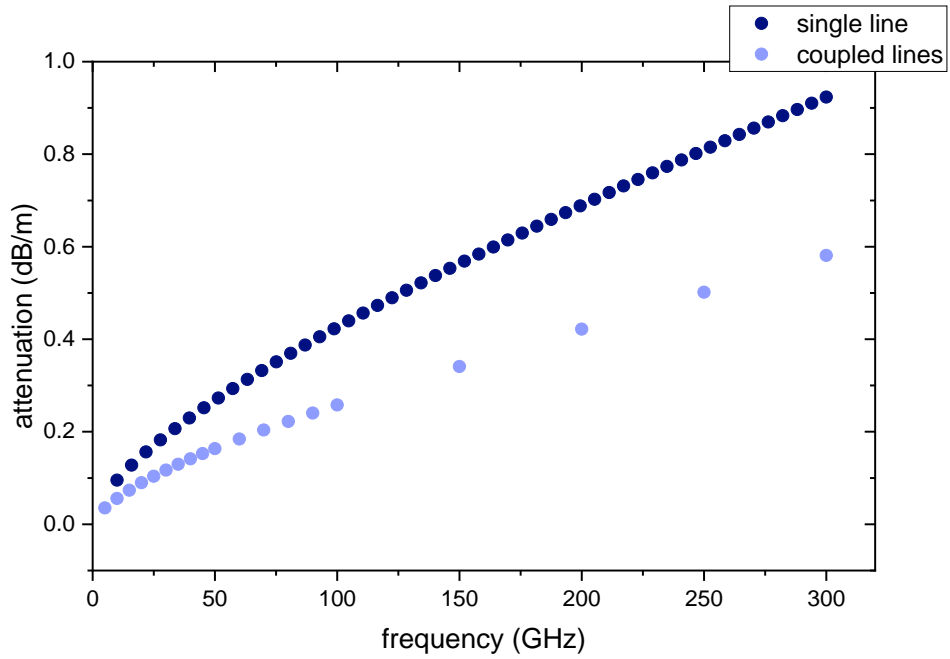

*Figure 14 Two Sommerfeld lines symmetric mode attenuation comparison with one line as function of frequency*

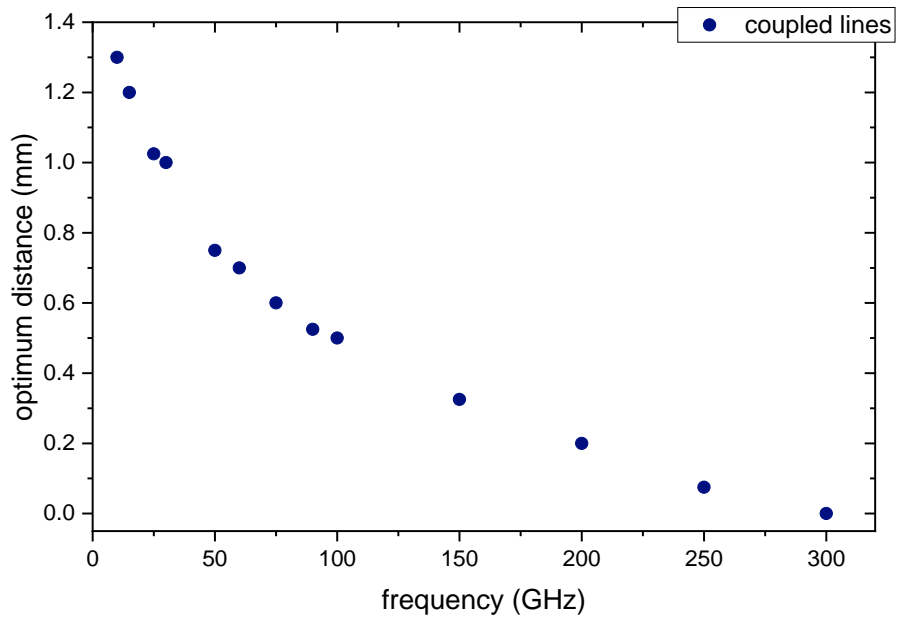

*Figure 15 Two coupled Goubau lines optimum d as function of frequency for symmetric mode*

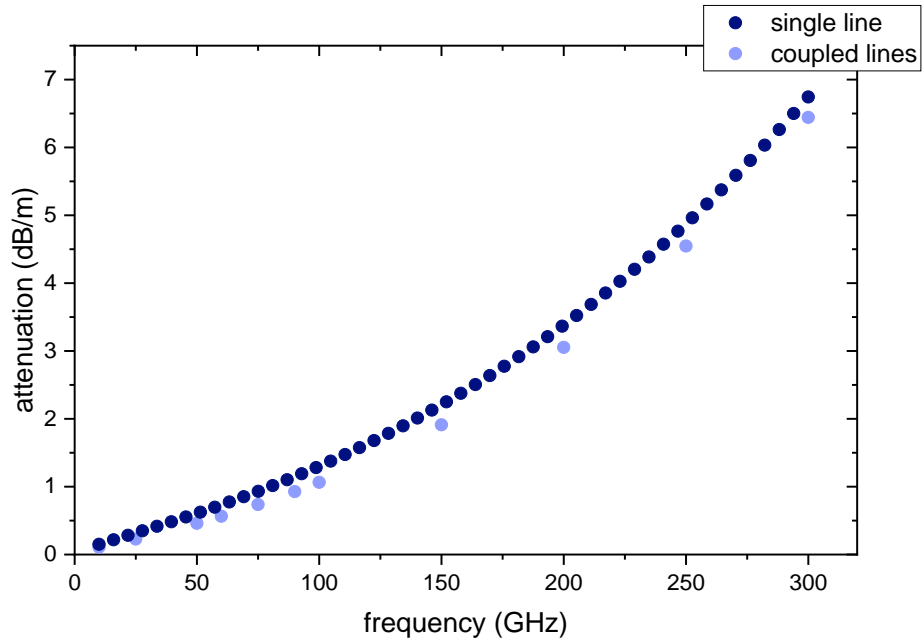

*Figure 16 Two Goubau lines symmetric mode attenuation comparison with single line as function of frequency*

Importantly, it can be seen from Figure 13 that in the Sommerfeld two wire case for the highest studied frequency the optimum separation distance for symmetrically coupled SWs is around 1.5mm (or 3 wire radii). However, for the Goubau line for the same symmetric mode the optimum distance is 0mm at 300GHz as can be seen in Figure 15. This holds significant practical importance since the twisted pair telephone wires are in direct contact. At these high frequencies, we expect the twist not to introduce a significant effect. However, it is known that SWs for single lines are sensitive to bends [26] [27], and it is true that according to our knowledge the effect of twist has not been thoroughly investigated yet. Nevertheless, we believe our results support the idea of using twisted pairs for SW transmission at these frequencies.

**Coupled wire waves in a dielectric bundle**

The arrangement studied above is an idealised one, as in reality the wires that are used in the telephone lines are surrounded by a bundle of other wires. In the case of a single surface wave wire this configuration results in increased longitudinal losses, due to the proximity of the perturbing dielectric close to the SW waveguide.

We consider the effect of these neighbouring wires for both the symmetric and antisymmetric modes. The configuration studied represents the one found in most telephone bundles, that is the two wires are used for SW transmission are surrounded by wires of the same cross section. In this model we assume that only the two innermost wires carry SWs, and investigate the propagation characteristics of both modes described above. Note that in this study we take a worst case scenario approach in which the other wires are completely filled with the same dielectric material as the coating of the two SW wires and they have no copper core.

The main characteristics of the attenuation curves are the same as for the case without the perturbing dielectric rods surrounding the wires(see Figure 10 and Figure 11). For the symmetric mode we found an optimum distance between the wires and the antisymmetric mode shows larger attenuation than the symmetric, as expected. At large separation distances the attenuation curves approach the single wire equivalent attenuation for both modes.

We found that in the case of two coupled wire waves the proximity of the dielectric bundle results in an increase of the longitudinal losses, about 0.1dB/m higher than without the perturbing dielectric bundle. This result is important for practical applications as it was thought before that one of the reasons why SWs are impractical to use in twisted pairs in telephone bundles, is the adverse effect of the surrounding wires on propagation losses.

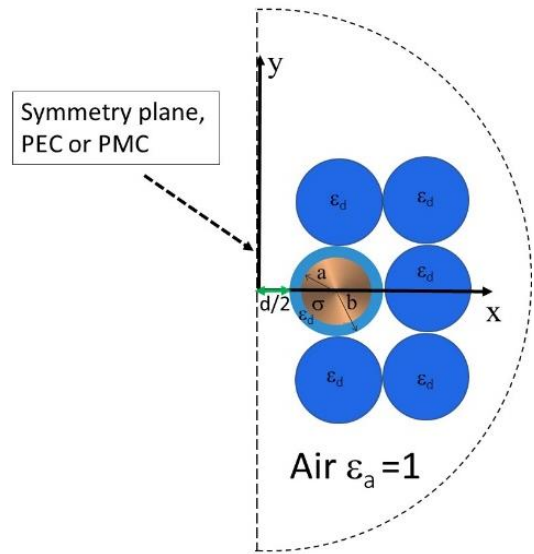

*Figure 17 Schematics illustration used to consider the effect of the perturbing dielectric bundle on two coupled, coated (Goubau) wires. The wires in the bundle all have the same dielectric properties as the coating of the SW carrying wires, but do not have the copper inner conductor.*

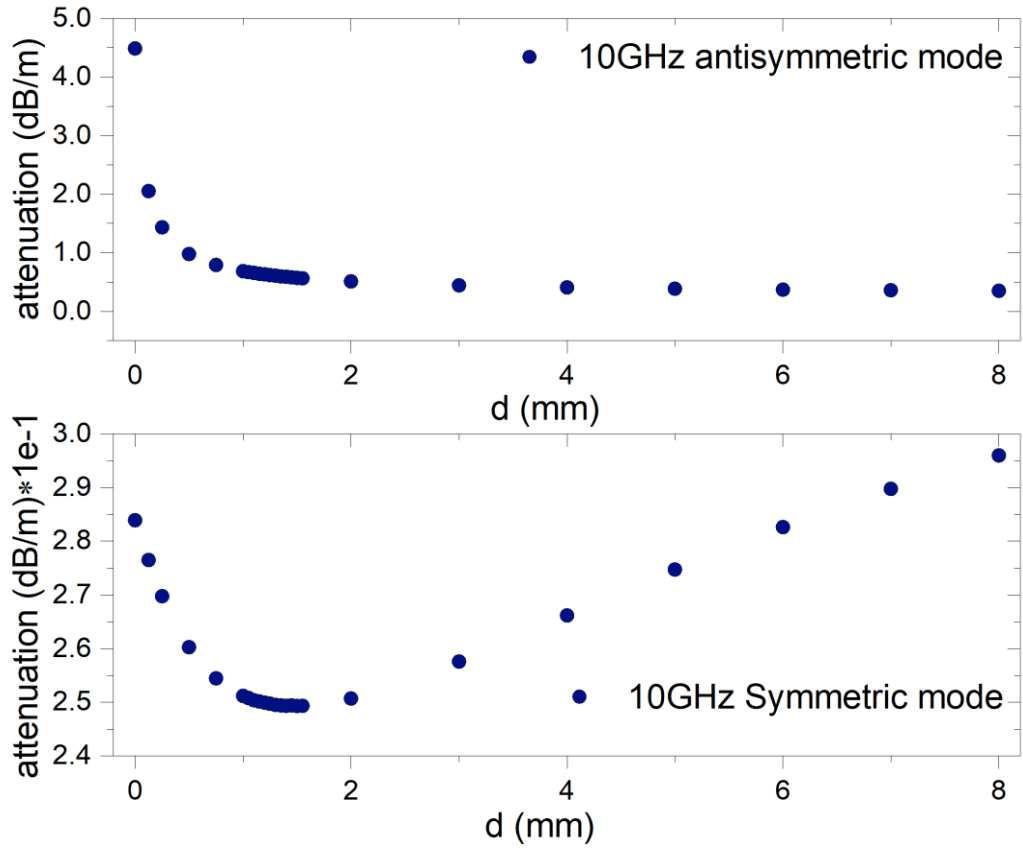

*Figure 18 Antisymmetric mode attenuation for two coupled Goubau lines in a bundle of wires (upper panel) and symmetric mode attenuation for two coupled Goubau lines in the same arrangement (lower panel) at 10GHz.*

**Experimental results**

Measurements have been carried out on surface waves on two wires to experimentally validate the theoretical results. We have measured the system as a two port device and have collected its S-parameters (cf Supplementary Material). The wires are both coated (Goubau line) and have identical material parameters to the ones studied in the preceding sections. However, they have slightly different geometrical parameters namely a=0.25mm and b=0.54mm (cf Figure 1). Importantly the thicker insulation and smaller metal cross section results in higher losses, thereby eliminating some practical issues with the noise floor. Nevertheless, we recalculated the attenuation for these parameters with the same methods as above for all separations and the comparison serves as validation of our theoretical results.

We are comparing calculated and measured $S_{21}$ parameters for different wire separations as a function of frequency in symmetrically coupled SWs guided by coated (Goubau) wires. Furthermore, in order to couple surface waves onto the wires specially designed planar launchers were used together with a splitter circuit to ensure that surface waves on each wire are in phase, and hence the symmetric eigenmode is measured (cf Supplementary Material). In the calculations we included the losses of the splitter circuit and launchers as well. Additionally, both the splitter circuit and the launchers have finite bandwidth in which their operation are optimized with respect to transmission (cf Supplementary Material). Taking this into account we present results between 2GHz and 6GHz in this work.

The results of both the measurements and calculations are plotted for a separation of 9mm and 25mm in Figure 19 and Figure 20 respectively. It is clear that there is excellent agreement between calculation and measurement results for the 9mm separation, the two curves follow the same qualitative shape and the difference is within a few dB in the whole frequency range. For the 25mm separation between the wires the same qualitative statement is true, however the noise is larger in this case. The agreement between theory and measurement is within less than a percent around 4GHz for both separations, where the launchers have their transmission peaks.

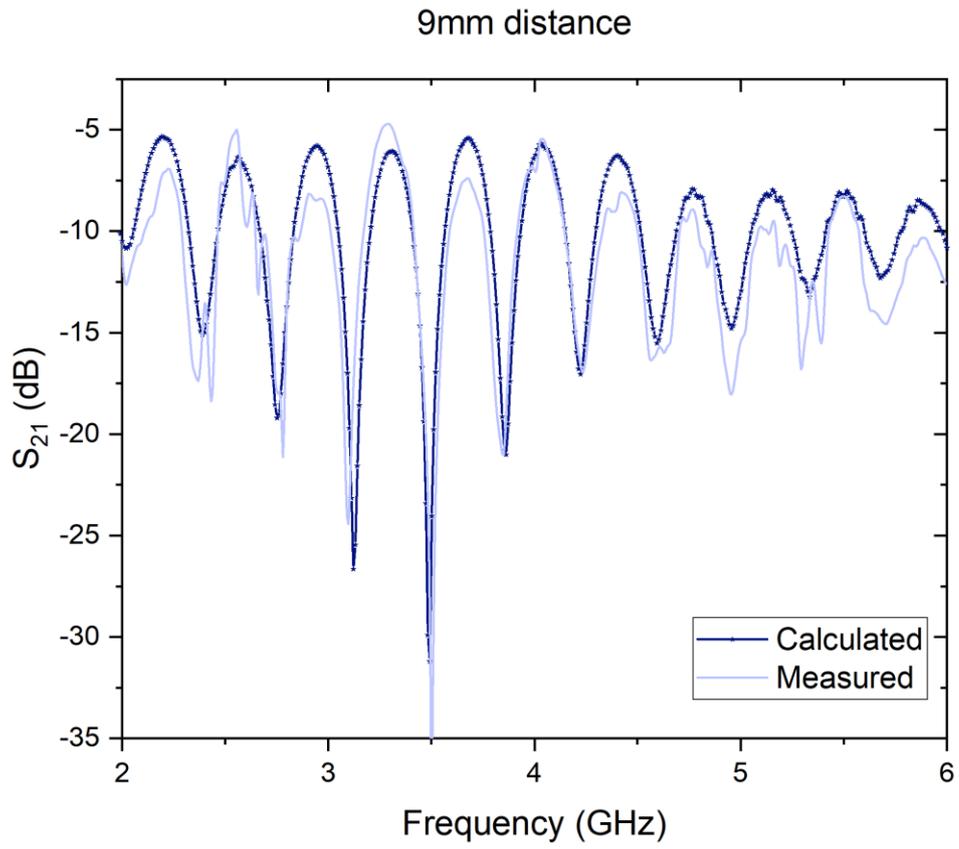

*Figure 19 Experimental(light blue line) and calculated (blue line and star) $S_{21}$ parameters of symmetrically coupled SWs on two Goubau wires for a separation of 9mm.*

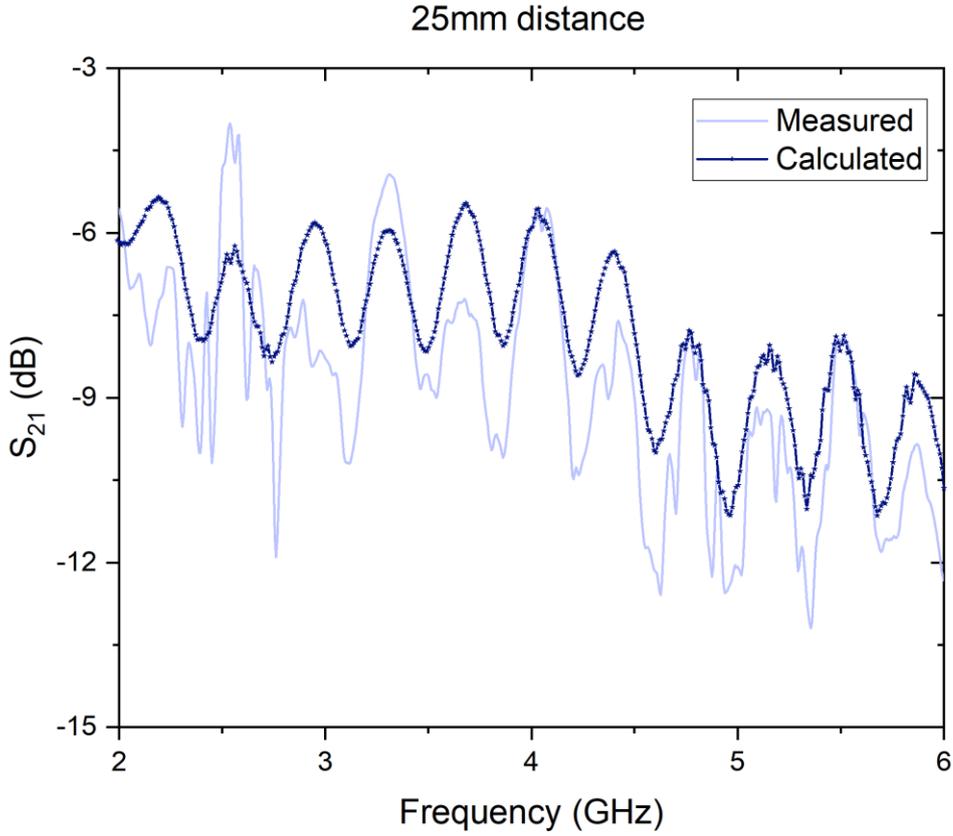

*Figure 20 Experimental(light blue line) and calculated (blue line and star) $S_{21}$ parameters of symmetrically coupled SWs on two Goubau wires for a separation of 25mm.*

**Discussion**

The differences between the analytic approach and the FEM numerical calculations originate from the assumptions taken in the former case (name it please). Importantly, at small separations the assumption that the field of the other wire at each wire's metal surface and the tangential electric field are negligible. The current density at each wire's surface can be written in the following form:

$$\vec{J}_T = \vec{J}_{SW} + \vec{J}_{IND} \qquad (5)$$

Where $\vec{J}_{SW}$ is the current density due to the SW on the wire and $\vec{J}_{IND}$ is the induced current by the other wire's fields. The loss in the conductor can be obtained by:

$$P_c = \oint \frac{1}{2} R_S |\vec{J}_T|^2 \, dl \qquad (6)$$

$J_{IND}$ has two components which comes from the fields $E_T$ and $H_\phi$. In the analytic approach, in the first order approximation $J_{IND}$ is neglected and therefore the power loss in the metal is independent of the distance between the wires, since it only depends on one wire's own current density. However, $H_\phi$. strongly depends on the distance between the wires as it is a function of radial coordinate as we have seen for the single wire (see Figure 3). As for $E_T$ its effect again is stronger at smaller distances, and it induces a current proportional to $E_T$ but with opposing signs on the two halves of the wire. At large distances this balances out, and the net field is close to 0 thus its effect can be neglected. At small distances this imbalance is greater and it has a non-zero contribution. $P_c$ can be obtained by numerical integration in COMSOL thus one can see how it depends on the relative distance for a given frequency. $P_c$ is plotted at 6GHz as a function of distance between the wires in Figure 21. Note, that the vertical line is the optimum distance between the wires, or where the upturn is observed in the attenuation coefficient. Since the attenuation coefficient behaves similarly for all frequencies we expect this to be a representative behaviour for all frequencies.

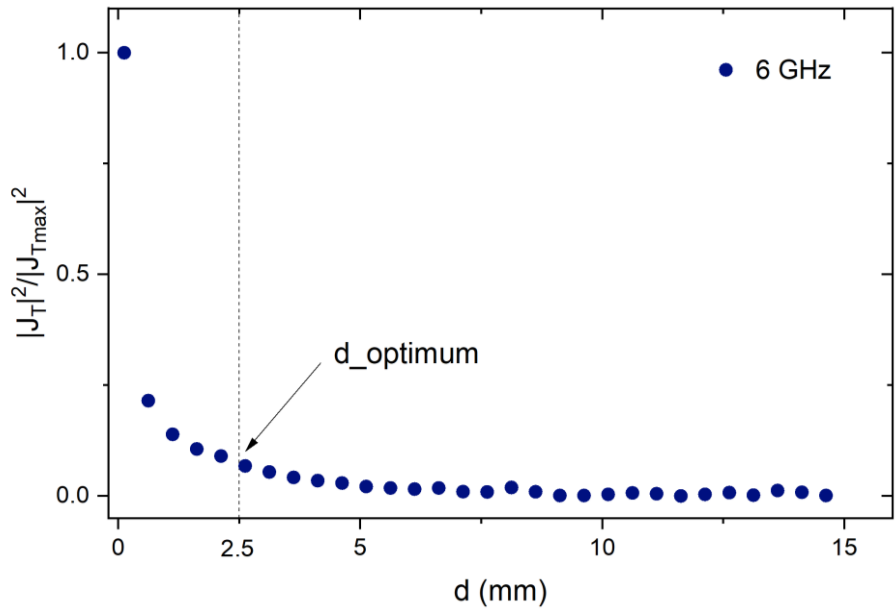

*Figure 21 The normalized total current density squared (resistive loss) on the metal surface as a function of distance between the wires for two symmetrically coupled Sommerfeld lines. It can be seen that $J_T$ is nearly constant for larger separations and only $J_{IND}$ can depend on the relative distance between the wires.*

In our approximation (based on Sommerfeld [8]) the transmitted power by the wave ($P_{TR}$) can be obtained by the following surface integral:

$$P_{TR} = \int_S (\vec{E}_1 \pm \vec{E}_2) \times (\vec{H}_1 \pm \vec{H}_2)\, dA \qquad (7)$$

The plus sign is for symmetric mode whereas the minus is for antisymmetric mode. The cross products between $E_i$ and $H_i$ are the single wire's Poynting vector, whereas the $E_j$ and $H_i$ ($\forall i$ & $j \in \{1,2\}$) are the two wire wave Poynting vector. The two wire terms are again dependent on the separation of the wires, the closer the wires the greater the coupling. Note, that this expansion of the fields is true in the weak approach, where the two-surface wave is a superposition of single surface waves. The attenuation is similarly obtained as in the single wire case equation (3):

$$\alpha = \frac{P_c}{2P_{TR}}$$

Therefore, the attenuation will depend on the relative magnitudes of the two opposing effects, on one hand smaller distance will result in greater induced currents in the other wire but on the other hand this will increase the power in the two-wire wave. It is clear from this explanation, why the weak coupling analytic approach misses the upturn in the attenuation in the case of symmetric coupling. It simply disregards the induced current parts at small distances and therefore $P_{TR}$ gets larger and larger whereas $P_C$ has no dependence on the relative distance between the wires.

This reasoning also holds true for the Goubau two wire wave, although the losses are more complicated, because they consist of the dielectric as well as metallic losses. The electric and magnetic fields have smaller radial extent at any given frequency resulting in a smaller optimum distance between the coated wires compared with the two Sommerfeld wires. This results in a sharper decrease of the optimum distance as a function of frequency and in the frequency range studied in this work the optimum distance between the coated wires is 0mm. It's worth mentioning that although Cook and Chu reported two coupled Goubau lines' effective mode indices as a function of separation for the lowest mode [24], their work can be also be considered in this sense weak coupling limit for this specific case. It is evident from the fact, that their method didn't produce an optimum separation distance for the symmetric mode. The existance of optimum distance is the result of strong coupling and we clearly obtain that with our numerical calculations.

The physical behaviour described above can also be understood from a simple intuitive picture. The attenuation will first decrease as d gets smaller, because a two-wire wave forms, which can be thought of as surface wave of the order of twice the effective cross section. However, as the wires get even closer in the symmetric mode the magnetic fields have opposite signs and this results in a magnetic field at the wire surfaces that creates an opposing induced magnetic field, thereby extinguishing the fields on the near sides of the wires (see Figure 22). This effectively reduces the metal cross section to which the field is 'bound', thereby increasing the losses, if the current is kept constant. In the analytic description however, we neglect the effect of the fields stemming from the other wire at each of the wires' metal surfaces. For large separation distances the fields resemble two non-interacting SWs on each wire (see Figure 24) in both the antisymmetric and symmetric cases, therefore the waves become uncoupled.

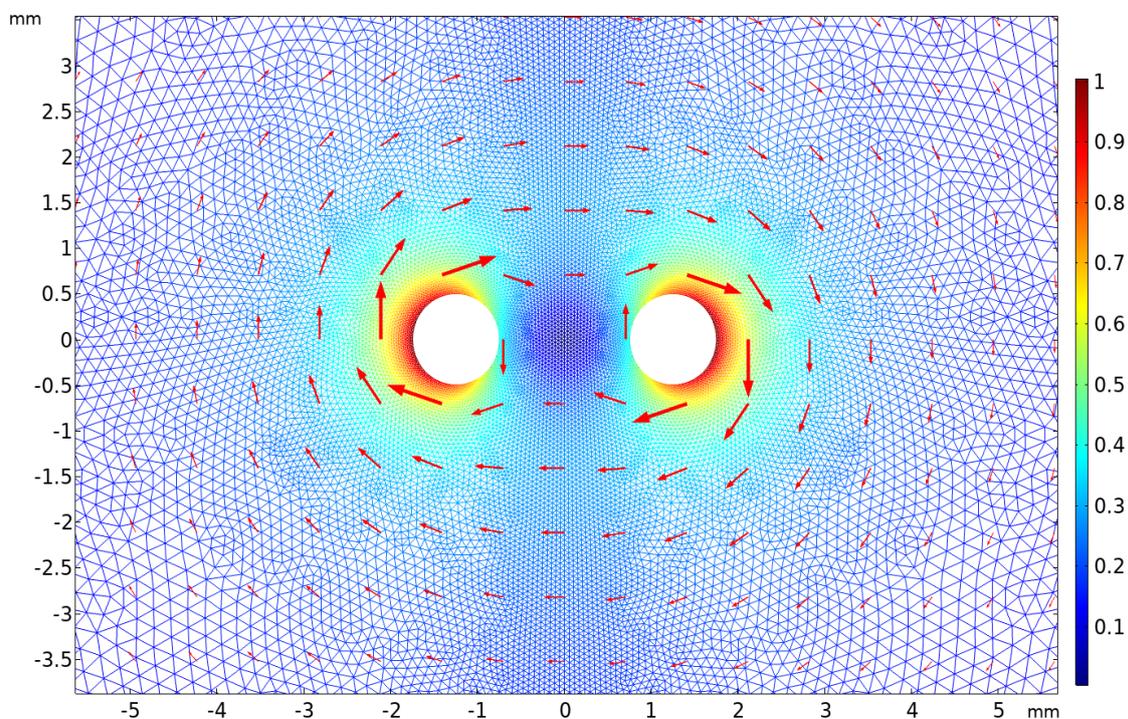

*Figure 22 Intensity of the total magnetic field for the symmetric mode at a small separation distance (d=1.5mm) between the wires at 10GHz in the Sommerfeld case. Note that only the outer part of the wires contributes to binding the wave in the two wave mode when they're close. The key is the*

*formation of the two wire wave, and at the optimum distance at which the extinction of the fields between the wires is not significant.*

For the antisymmetric mode we expect the attenuation to increase as the separation of the wires is decreased for both the analytic and the FEM numerical models, as in this mode, the two wire SW cannot form even for small separations because of the opposing signs of the SWs on each conductor. The field profile will always resemble that of two separate SWs at each wire. Therefore as we gradually decrease the separation between the wires the attenuation will get higher and higher in both the analytic and FEM models. However, again since we neglect the fields from the other wire at each of the wires' surfaces at small separations the fields strongly interact and couple in a manner that the effective resistance of the wires increases. In Figure 23 a small separation distance is shown and it can be seen that the fields are concentrated between the two wires, therefore the fields only bind to a fraction of the inner sides of the wires. At large distances the field profile will again resemble two non-interacting SWs on each wire (see Figure 25). Therefore, the attenuation will asymptotically approach the attenuation of the single wire SW from above.

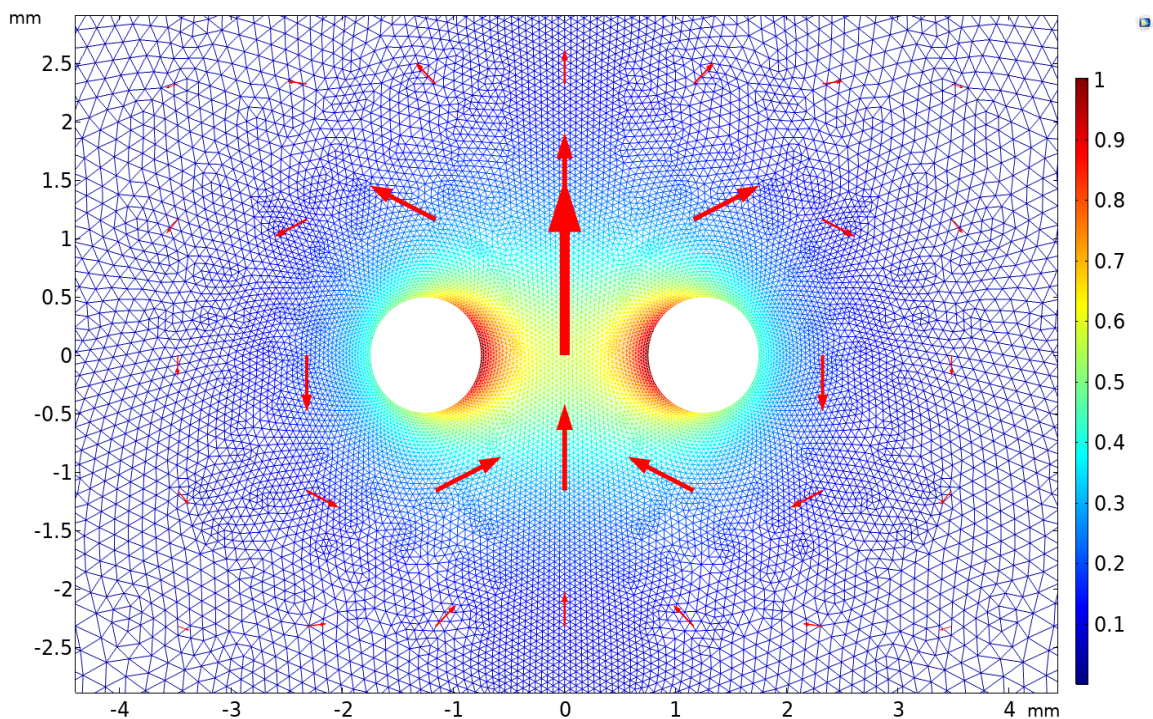

*Figure 23 Intensity of the total magnetic field for the antisymmetric mode, at small seperation distance ( d=1.5mm) for two Sommerfeld wires. Note that at this seperation the field is concentrated between the wires, therefore the effective surface area of the wire is greatly reduced compared to a single wire case. In the antisymmetric mode even in close range there is not two conductor surface wave scenario, rather the field profile resembles two single wire surface waves but with smaller cross section because of the reduced intensity of the field on the outer perimeter of the wires.*

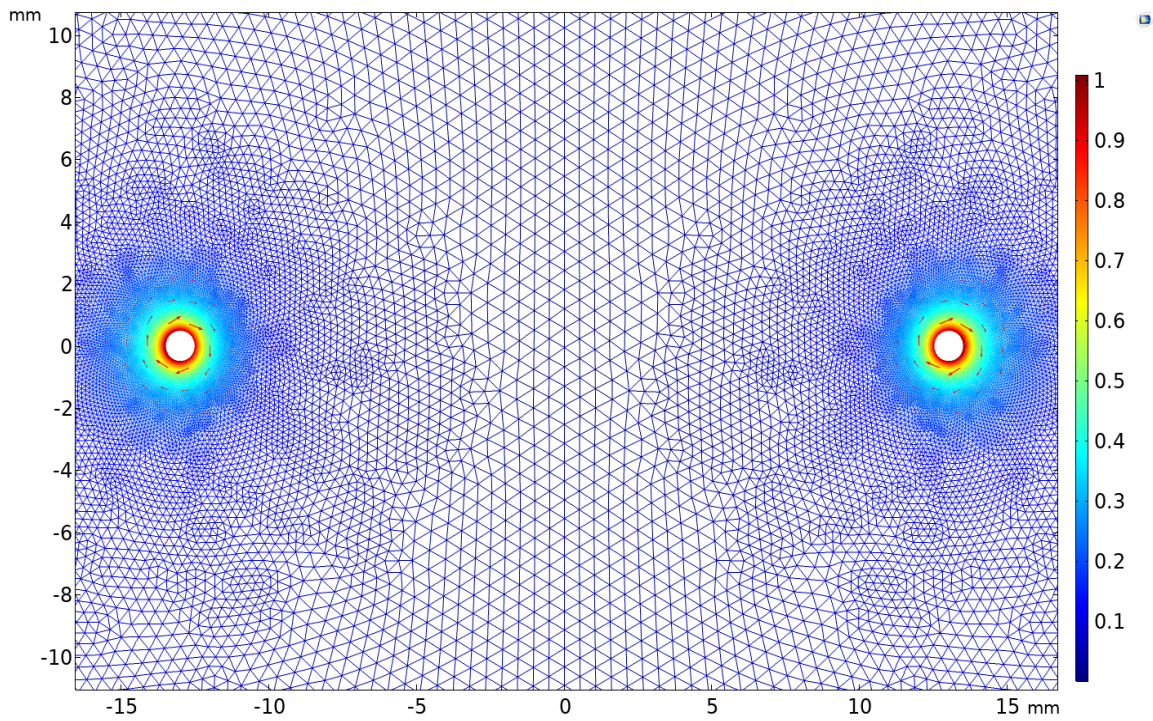

*Figure 24 Intensity of the total magnetic field for the symmetric mode at large separation distance (d=29mm) between the wires at 10GHz for two Sommerfeld wires. At large seperation distances between the wires the field profile looks as if two independent one wire waves on each wire are propagated out of the plane in the z direction.*

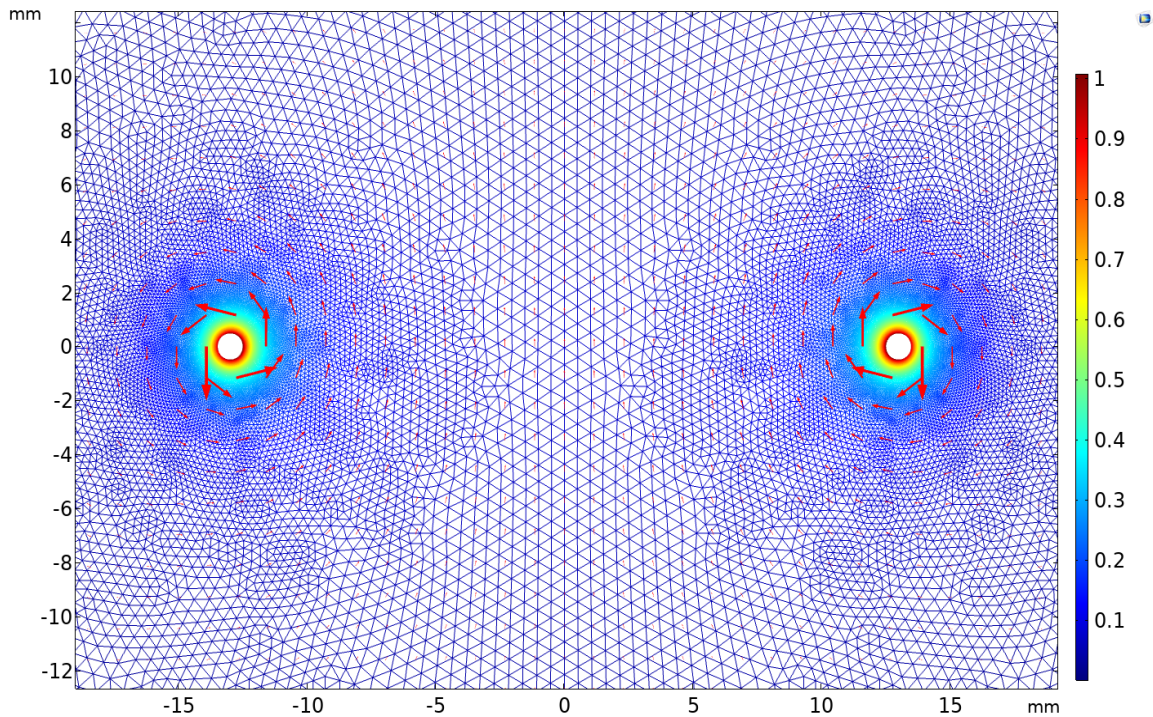

*Figure 25 Intensity of the total magnetic field for the antisymmetric mode at large separation distance (d=29mm) between the wires at 10GHz for two Sommerfeld wires. At large seperation distances between the wires the field profile looks as if two independent one wire waves on each wire are propagated out of the plane in z direction.Also note, however, that this is only true exactly in the far limit and there's a slight assymmetry still present in the field profile on the wire cross section at finite distances.*

As we have shown, for large separation distances both modes' attenuation ought to approach the attenuation of the single wire SW at any given frequency. As expected, the antisymmetric mode will asymptotically approach it from above and the symmetric mode from below as the function separation distance. At 100 GHz this can be clearly seen from the attenuation plot in Figure 26.

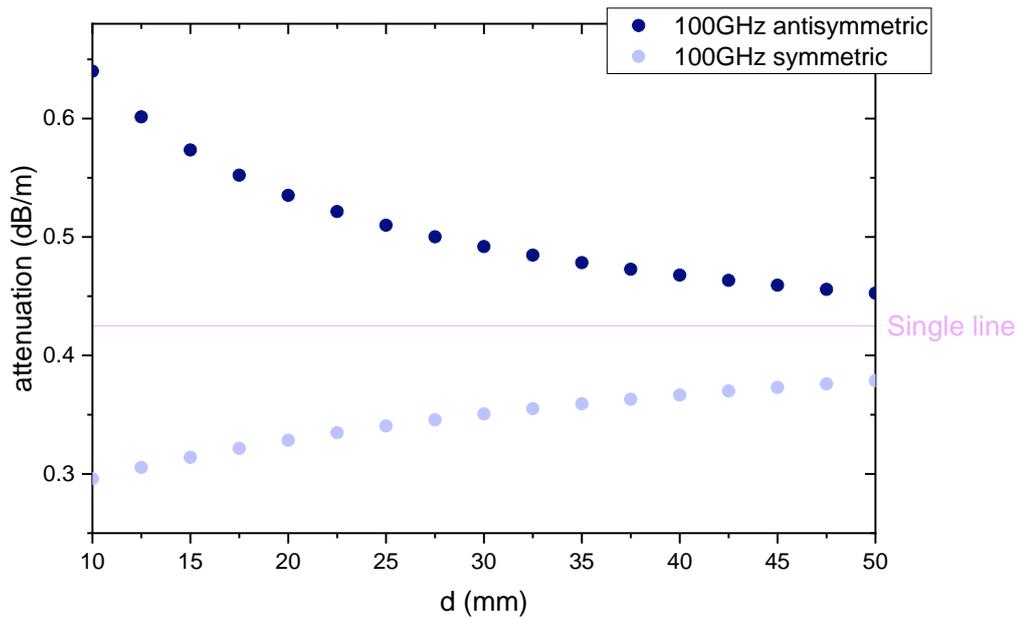

*Figure 26 Attenuation coefficient for two coupled Sommerfeld lines for both symmetric and antisymmetric mode at 100GHz for large separations. Horizontal line is the standalone line equivalent attenuation at 100GHz. Note that both the symmetric and asymmetric attenuation curves asympotically converge to the attenuation of a single line. At higher frequencies this occurs at smaller distances, because of the more tightly confined fields around the wires.*

The intuitive picture presented above also applies to the coupled Goubau wire waves, however there are slight modifications. The wires are coated and therefore the metal surfaces cannot get arbitrarily close to each other, which would result in very large losses. It is instructive to think of the Gobau line as a Sommerfeld line surrounded by a different medium than air. Then at very high frequencies the single wire wave's propagation constant will approach the propagation constant of the dielectric as the SW becomes very closely bound to the dielectric. However, before that regime is reached in our two wire SW case the wires' coating could touch each other, at which point the separation between the wires 0mm. From our calculations we conclude that at 300GHz the optimum separation of the coated wires is 0mm.

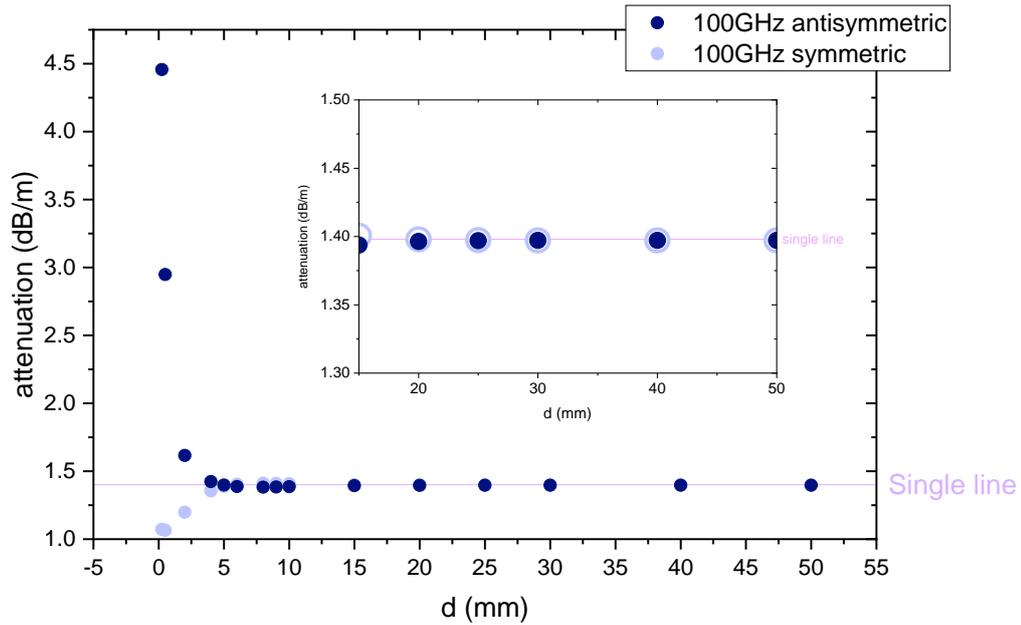

*Figure 27 Large d behaviour of symmetric and antisymmetric modes at 100GHz of two coupled Goubau lines. Horizontal line is the standalone line equivalent attenuation at 100GHz. Note that both the symmetric and asymmetric attenuation curves asympotically converge to the single line attenuation. In the case of Goubau lines the fields are more tightly confined compared to the Sommerfeld case at any given frequency and they converge to the single line at smaller distances.*

**Conclusion**

In this paper, we have developed FEM-based and analytical models to uncover unexplored properties coupled surface wave lines. The models considered both Sommerfeld and Goubau lines to investigate coupling in dual uncoated and coated lines, respectively. Both models have been validated against each other for the two Sommerfeld lines scenario, providing remarkable agreement within 1% except for symmetrically coupled wires over extremely small separation distances (or in terms of free space wavelength $\lambda_0$ and wire radius a, $\lambda_0/a>4.2e-3$). Results suggest that antisymmetric surface wave can suffer attenuation losses as high as 3 dB/m whilst symmetrically coupled surface waves exhibit significantly lower attenuation 0.064 dB/m in the Sommerfeld case at 0mm separation which is approximately 46 fold reduction in attenuation loss. The effects of coupling in the Goubau case are less dramatic than latter with values as high as 0.5 dB/m in the antisymmetric scenario and 0.11 dB/m in the symmetric case at 1mm separation, i.e. 4.5 fold reduction in loss. This is due to the fact that the fields are more confined laterally and do not expand as far as in the Sommerfeld scenario. The results, both via FEM-based and analytical models, additionally demonstrated the convergence of the attenuation loss to the single line case when coupling occurred over large separation distances. We have also validated our results by means of experiments on coupled SWs. The measured and calculated results show remarkable agreement at 4GHz within 1% of each other. The findings of this paper signify the nature and the role of coupling in determining the losses of surface wave mode in multi-conductor environments particularly for both short and long range communication systems. These results are novel and the mode of coupling have been not investigated to the authors' best knowledge, and particularly the frequency dependence have not been mapped. The short range effects, particularly the upturn of the attenuation of symmetric waves, uncovered by the numerical results is an entirely novel finding of this paper for coupled SWs.

**Acknowledgements**

This work was supported by the Royal Society Grants IF170002 and INF-PHD-180021. Additional funds were provided by BT plc and Huawei Technologies Co., Ltd. The authors thank the Royal Society, BT and Huawei for these funds.

Supplementary Material for

# Interaction between Surface Waves on Wire Lines

D.Molnar[1]*, T. Schaich[1], A. Al Rawi[1,2] and M.C. Payne[1]

[1]Theory of Condensed Matter Group, Cavendish Laboratory, University of Cambridge, CB3 0HE, UK

[2]BT Labs, Adastral Park, Orion building, Martlesham Heath, IP5 3RE, UK

*corresponding author, dm611@cam.ac.uk

**S1 Derivation of weak coupling limit, analytical approach of the characteristic equation**

In this section we derive the characteristic equation to find the complex propagation constant ($\beta$) of the coupled wire wave. Note that for obvious reasons we will refer to this approach as analytic but in reality in the last step we made use of a numerical solution of the coupled characteristic equation, just as we did when we solved for the single wire.

We use two cylindrical coordinate systems (r; $\phi$; z) and (r'; $\phi$'; z') with their respective centres in the middle of the conductors and their z-axes parallel. The fields are assumed to be a superposition of the single wire surface wave outside and a standard single wire SW inside the conductors. The conductors are assumed to be made of the same material and have radius a. A common factor of $e^{i(\omega t - \beta z)}$ suppressed in this derivation. We split the electric field into a transverse field $\vec{E}_T$ and a longitudinal field $\vec{E}_z$ such that the total field $\vec{E}$ is given as $\vec{E} = \vec{E}_T + E_z \cdot \hat{z}$, where $\hat{z}$ is the unit vector in the z-direction. For the magnetic field $\vec{H}$, we can express the fields outside the conductors as in the following:

$$\vec{E}_z = B_1 H_0(\gamma r)\vec{e}_z + B_2 H_0(\gamma r')\vec{e}_{z'} \qquad 1.1$$

$$\vec{H}_\phi = \frac{i\omega\varepsilon}{\gamma}\left(B_1 H_1(\gamma r)\vec{e}_\phi + B_2 H_1(\gamma r')\vec{e}_{\phi'}\right) \qquad 1.2$$

$$\vec{E}_T = \frac{i\beta}{\gamma}\left(B_1 H_1(\gamma r)\vec{e}_r + B_2 H_1(\gamma r')\vec{e}_{r'}\right) \qquad 1.3$$

$$\gamma = \sqrt{\omega^2\varepsilon\mu - \beta^2} \qquad 1.4$$

where $\varepsilon$ and $\mu$ are the permittivity and permeability of the surrounding medium, B1,2 are constants. Inside the conductors we have:

$$\vec{E}_z = A_1 J_0(\gamma_c r)\vec{e}_z + A_2 J_0(\gamma_c r')\vec{e}_{z'} \qquad 1.5$$

$$\vec{H}_\phi = A_1 \frac{i\omega\varepsilon_c}{\gamma_c} J_1(\gamma_c r)\vec{e}_\phi + A_2 \frac{i\omega\varepsilon_c}{\gamma_c} J_1(\gamma_c r')\vec{e}_{\phi'} \qquad 1.6$$

$$\vec{E}_T = A_1 \frac{i\beta}{\gamma_c} J_1(\gamma_c r)\vec{e}_r + A_2 \frac{i\beta}{\gamma_c} J_1(\gamma_c r')\vec{e}_{r'} \qquad 1.7$$

$$\gamma_c = \sqrt{\omega^2\varepsilon_c\mu_c - \beta^2} \qquad 1.8$$

where $\varepsilon_c$ and $\mu_c$ are the permittivity and permeability of the conductor respectively. $H_0, H_1, J_0$ and $J_1$ stand for the Hankel and Bessel functions. Let d be the distance between the surfaces of our cylinders. Then, d + 2a is the distance from the centres of our coordinate systems. We assume that the conductors are small such that the fields originating from one conductor do not vary much over the other one. In addition, if d $\gg$ a these fields will be small. Hence, we can formulate approximate boundary conditions by only considering the perturbing effect of the other conductor in the longitudinal electric field. This gives

$$B_1 H_0(\gamma a) + B_2 H_0\big(\gamma(d + 2a)\big) = A_1 J_0(\gamma_c a) \qquad 1.9$$

$$B_1 H_0\big(\gamma(d + 2a)\big) + B_2 H_0(\gamma a) = A_2 J_0(\gamma_c a) \qquad 1.10$$

$$B_1 \frac{i\omega\varepsilon}{\gamma} H_1(\gamma a) = A_1 \frac{i\omega\varepsilon_c}{\gamma_c} J_1(\gamma_c a) \qquad 1.11$$

$$B_2 \frac{i\omega\varepsilon}{\gamma} H_1(\gamma a) = A_2 \frac{i\omega\varepsilon_c}{\gamma_c} J_1(\gamma_c a) \qquad 1.12$$

Note that we only take into account the longitudinal electric field which is a first order approximation. To solve these equations, first we find the condition
$$B_1^2 = B_2^2$$
Without loss of generality we may choose $B_1$ = 1. Then we have the two options $B_2$ = 1 and $B_2$ = -1 corresponding to a symmetric and anti-symmetric mode. It is easy to show that for the symmetric case $A_1 = A_2$ and for the anti-symmetric case $A_1 = -A_2$ from the above equations.

The characteristic equation for the two modes is then found to be

$$\frac{\gamma}{\varepsilon} \frac{H_0(\gamma a) \pm H_0\big(\gamma(d + 2a)\big)}{H_1(\gamma a)} = \frac{\gamma_c}{\varepsilon_c} \frac{J_0(\gamma_c a)}{J_1(\gamma_c a)}$$

*Supplementary Equation 1.13*

The upper sign corresponds to the symmetric, the lower to the anti-symmetric mode. As in the case for the single Sommerfeld wire we can make use of the large argument expansion for the Bessel function to find $\frac{J_0(\gamma_c a)}{J_1(\gamma_c a)} \sim i$. The transcendental equation is then solved with standard numerical methods by using Matlab, and hence the lateral complex propagation constant $\gamma$ and from it the longitudinal complex propagation constant ($\beta$) can be obtained.

**S2 Numerical model**

To model SW propagation on a single wire and later on coupled wires we utilize the commercial FEM code COMSOL Multiphysics. The frequency domain Helmholtz equation is solved, where the $e^{i\omega t}$ dependence is assumed throughout. By solving this equation the out-of-plane component of the electric field is obtained and the rest of the field components are obtained from Maxwell's equations. We obtain the longitudinal complex propagation constant ($\beta$), of the out-of-plane z direction, which has real part $\nu$, the propagation constant, and complex part, $\alpha$ the longitudinal attenuation.

$$\vec{\nabla} \times \mu_r^{-1}(\vec{\nabla} \times \vec{E}) - k_0^2 \left(\varepsilon_r - \frac{i\sigma}{\varepsilon_0 \omega}\right)\vec{E} = \vec{0}$$

*Supplementary Equation 2. 1*

Helmholtz Vector Equation solved in COMSOL Multiphysics

$\mu_r$ is the relative magnetic permittivity and is assumed to be 1 for all materials throughout.

**Domain and Boundary conditions**

The boundary conditions applied when solving *the Helmholtz* equation in the FEM approach are the following (see Supplementary Figure 1). On the outer boundary if R, the radius of the circular air region surrounding the wire is chosen large enough it could either be a perfect electric conductor (PEC, $\vec{n} \times \vec{E} = 0$) or perfect magnetic conductor(PMC, $\vec{n} \times \vec{H} = 0$).

However, in practice R is limited by the computational resources and a low reflecting boundary condition or a perfectly matched layer (PML) can be chosen to mimic infinite space. In this work we chose the latter, and in the following the outer most boundary of the PML is chosen to be PEC, but PMC could equivalently suffice.

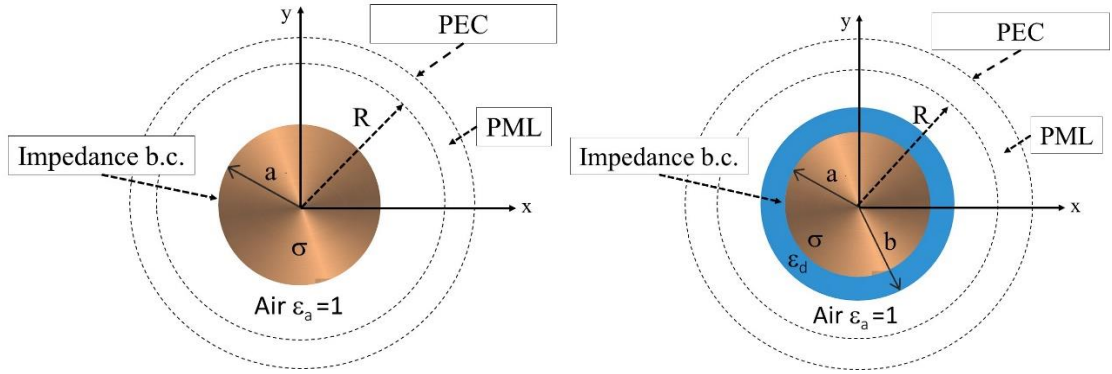

**Supplementary Figure 1:** Cross section schematics to show the boundary conditions used for single wires in this work.

$$\sqrt{\frac{\mu_0 \mu_r}{\varepsilon_0 \varepsilon_r - \frac{j\sigma}{\omega}}} \vec{n} \times \vec{H} + \vec{E} - (\vec{n} \cdot \vec{E})\vec{n} = 0$$

*Supplementary Equation 2. 2*

Impedance boundary condition (IBC)

It is essential for the existence of the Sommerfeld type wave that the conductor has a finite resistivity for the wave to have a smaller phase velocity than the corresponding phase velocity in air. For the Goubau wire wave it is not the case anymore, since it was Goubau's discovery that this condition can be relaxed if the metal wire is coated with a dielectric. Nevertheless, for more realistic results the finite conductivity is important in this case as well. At the GHz range the skin depth of good conductors, and particularly that of copper is on the order of 0.1µm. In our case that is 3 orders of magnitude smaller than the dimensions of the wire itself. Therefore, the fields practically only penetrate the wire to this depth and it can be assumed that in this case the metal can be replaced with a boundary without having to mesh perpendicular to it and resolve the skin depth. In COMSOL this boundary condition is called the impedance boundary condition (IBC), which allows for finite conductivity and surface currents to be computed. This condition is utilized to model the metal wires (copper wire) at these frequencies throughout this work.

The mesh element size is chosen according to the Shannon-Nyquist sampling condition as well as to resolve the smallest geometry features in the model. In our case using second order mesh element shape functions, it translates to a maximum element size of $\lambda/6$, where $\lambda$ is the free space wavelength. At 10GHz it is around 5mm and at 300GHz it is about 0.1mm. However, since the wire shape needs to be resolved accurately the smallest size is ultimately set by the geometry constraint in our case, that is the wire radius of 0.5mm or more precisely a fraction of that. We found that by having 15 mesh elements for each quadrant of the circular cross section of the wire, therefore having 60 elements in total around the wire's perimeter, the resolution is sufficient. As it can be seen in Supplementary Figure 2, COMSOL's meshing algorithm automatically grows the element size from this to the required lam/6 in the uniform, air domain surrounding the wire, but this growth rate can be made fully manually controlled. For the PML, because of the underlying coordinate stretching formula, a special mesh type a mapped mesh is used. Otherwise standard triangular elements suffice throughout the model. The colouring scheme is according to the mesh element size of the sides of the triangle (or quadraliteral elements in the PML) measured in m. The mesh is very similar for the Goubau wire case, however we also need to resolve the thickness of the coating. Again this is satisfied by using 15 elements per quadrant on both sides of the coating (metal and air side).

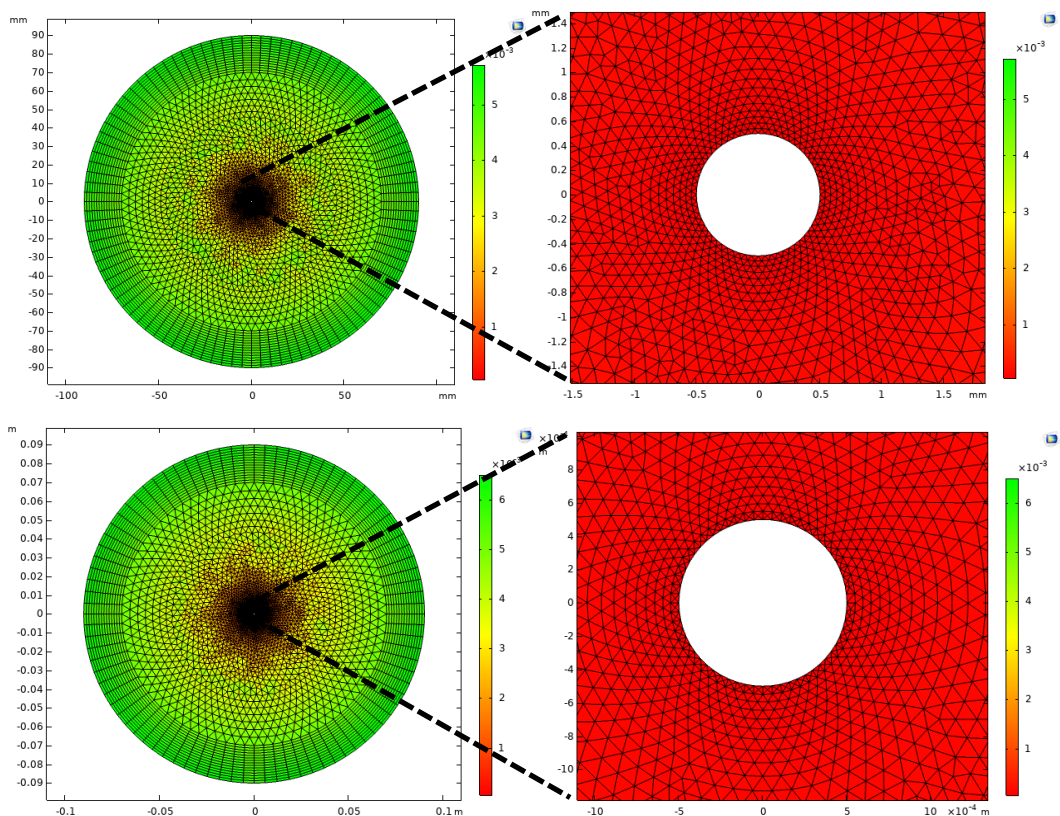

**Supplementary Figure 2:** Mesh for this present model. The elements are coloured by the mesh size according to the scale in meters. Upper graph is the Sommerfeld wire case and below is the Goubau wire. The enlarged region shows the mesh in the vicinity of the wires.

**Coupling FEM Boundary conditions**

Following the analytic approach for the coupling presented in the previous section we set out to investigate the same configuration numerically, with the use of FEM. The equation we are solving in this case is identical to the Helmholtz equation of the single wire. The boundary conditions are identical to the ones used in the standalone wire wave cases (see Supplementary Figure 1), however now we utilize the symmetry inherent in the geometry for the two wire wave case.

Only one half of the geometry is modelled directly, meshed and subsequently the *Helmholtz equation* is only solved over this half domain. The full geometry results, for instance field plots can be recovered in post-processing. This results in a great reduction of the number of degrees of freedom, thereby reducing the required computational resources and solution time. On the symmetry plane therefore we require either PMC or PEC boundary condition. The PEC corresponds to the antisymmetric mode whereas the PMC corresponds to the symmetric mode [25].

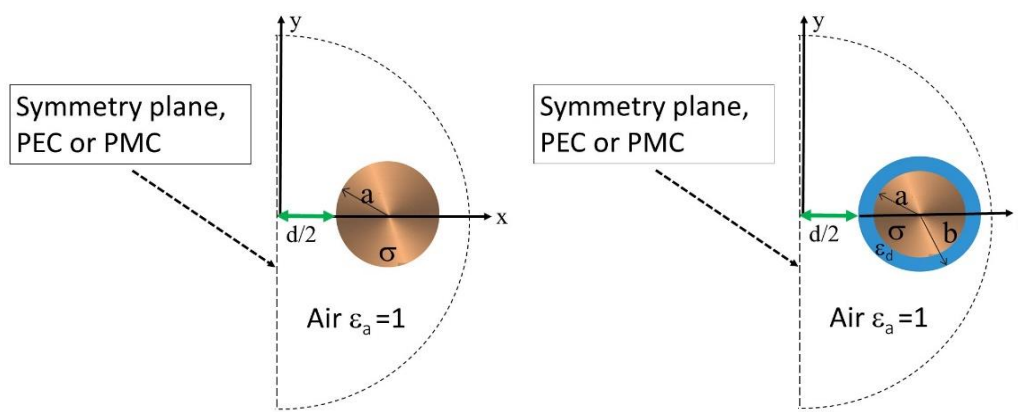

**Supplementary Figure 3:** Schematics and boundary condition of the coupled SW for both Sommerfeld and Goubau wire waves. For the antisymmetric mode the b.c. on the mid-plane is PEC whereas for symmetric mode it is PMC. The other boundary conditions are identical to that of single wires, except for the symmetry plane between the wires (see Supplementary Figure 1).

**S3 Experimental methods**

We utilized standard scattering parameter (S-parameter) measurements. We used a 2 channel, 8722ET Transmission/Reflection Vector Network Analyzer (VNA), with a range of 50 MHz to 10 GHz, with 1601 sampling points in the frequency interval. The schematic representation of the measurements can be seen in Supplementary Figure 4. The distance between the two SW carrying wires (SW line1 and SW line2) are varied and the S-parameter matrix of the system is measured and recorded with the VNA. The coaxial cables connecting the ports of the VNA to the splitters are calibrated out during the calibration process. For the SW mode excitation planar launchers were used, based on THz plasmon launchers by Akalin et al. [28]. By applying scaling theory, the dimensions of the launchers are scaled up for GHz applications. The launchers were designed and manufactured in our in-house facilities.

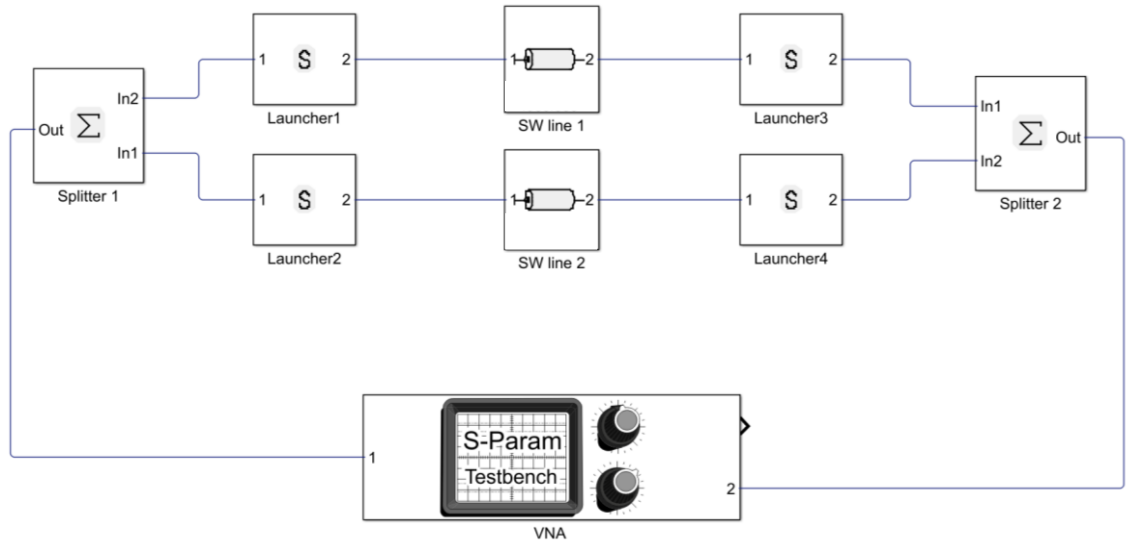

**Supplementary Figure 4** Schematic diagram of the measurement setup

The actual measurement setup with the SW launchers on the emitter side of the wires is shown in Supplementary Figure 5. The receiver side of the wires is identical to the emitter. The support jig has slots for the launchers on both emitter and receiver side, facilitating controlled variation of the separation of the wires.

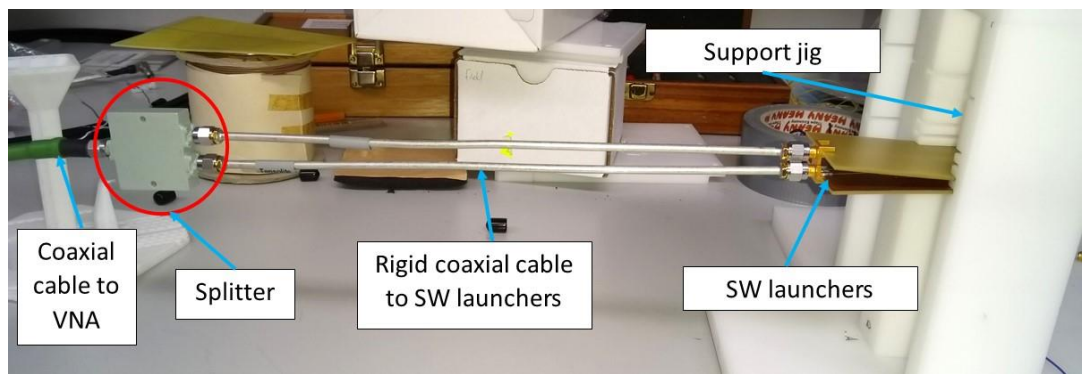

**Supplementary Figure 5** picture of the measurement of the S parameters of coupled SWs

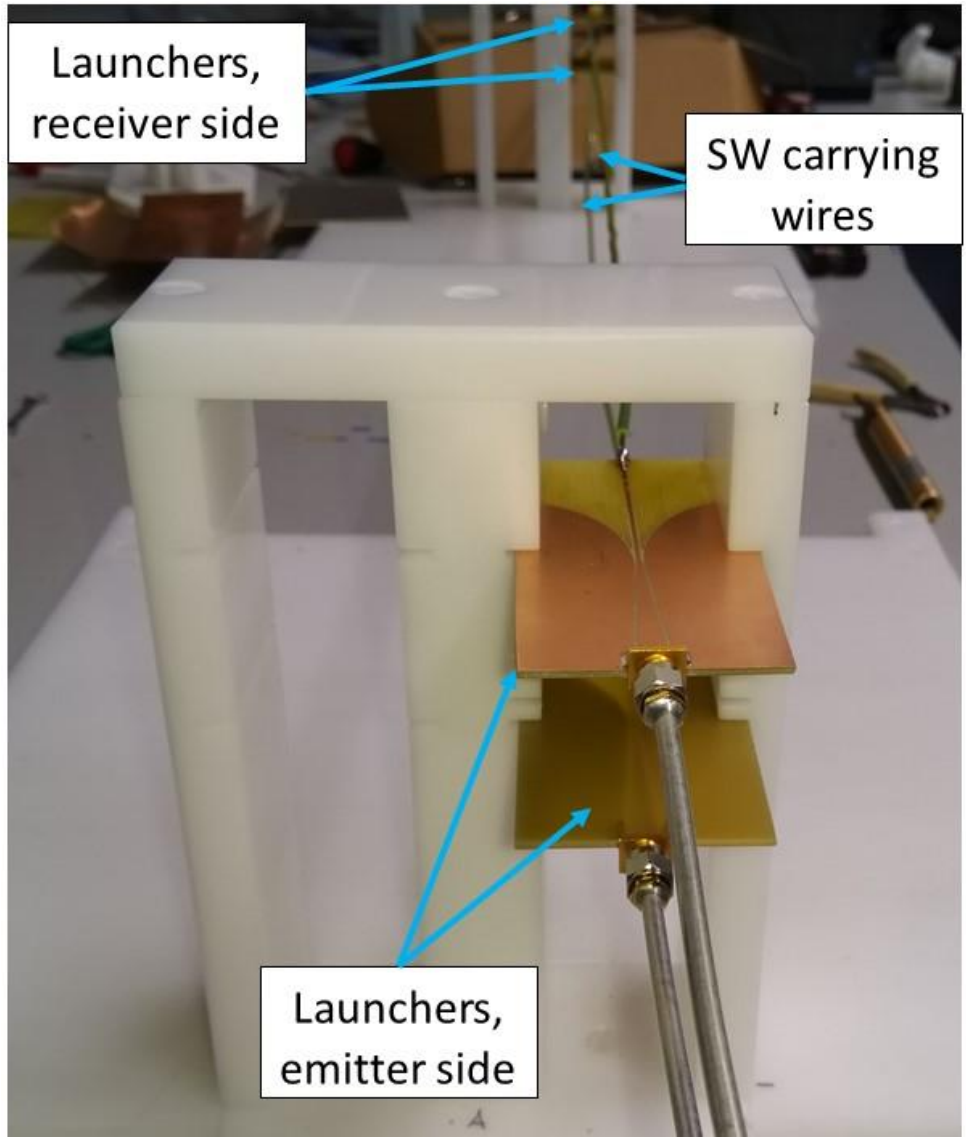

**Supplementary Figure 6** Close -up picture of a pair of launchers inserted into the support jig and experimental setup.

In order to compare our calculations on the SW coupling, the additional components of the measurement setup were measured separately. The S-parameters of the splitters and the coaxial cables used to connect them to the launchers were also separately recorded according to Supplementary Figure 7.

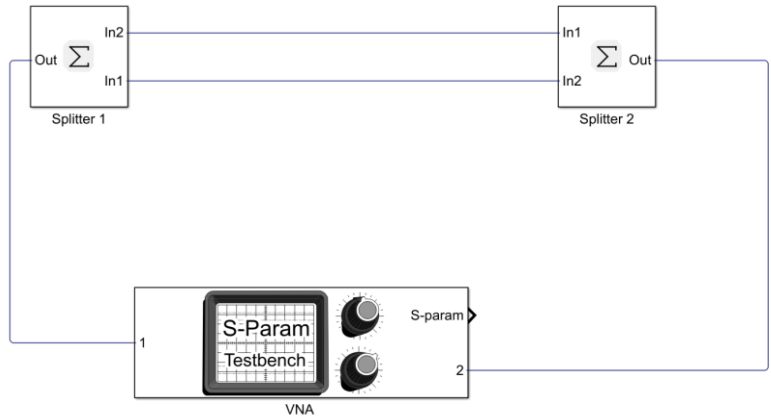

**Supplementary Figure 7** Schematic diagram of the splitter measurement

The S parameters of the launchers was also measured in a back-to-back configuration in which two planar launchers are connected to each other without the wires. The schematic diagram of the measurement can be seen in Supplementary Figure 8. These S parameters were added to the calculated two wire losses and compared to measured $S_{21}$ values of the two wire Goubau waves. The $S_{21}$ measurement data of the back-to-back launchers and the splitter circuits are shown in Supplementary Figure 9. The green lines indicate the frequency range in which launchers and splitters are best performing in terms of their transmission characteristics. The launchers have a peak transmission at 4GHz, at which frequency we obtain perfect match between the calculated and measured results.

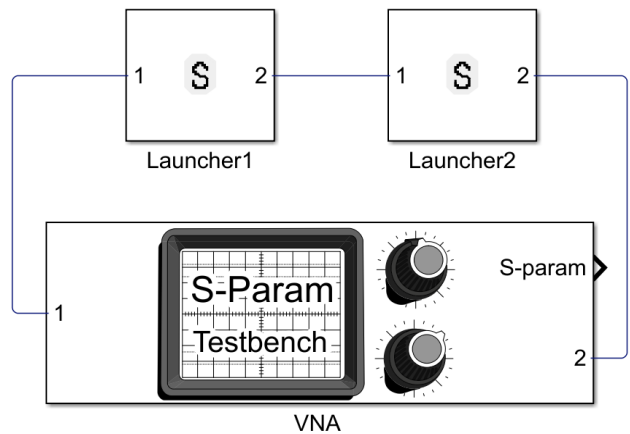

**Supplementary Figure 8** Schematic diagram of the surface wave launcher measurement

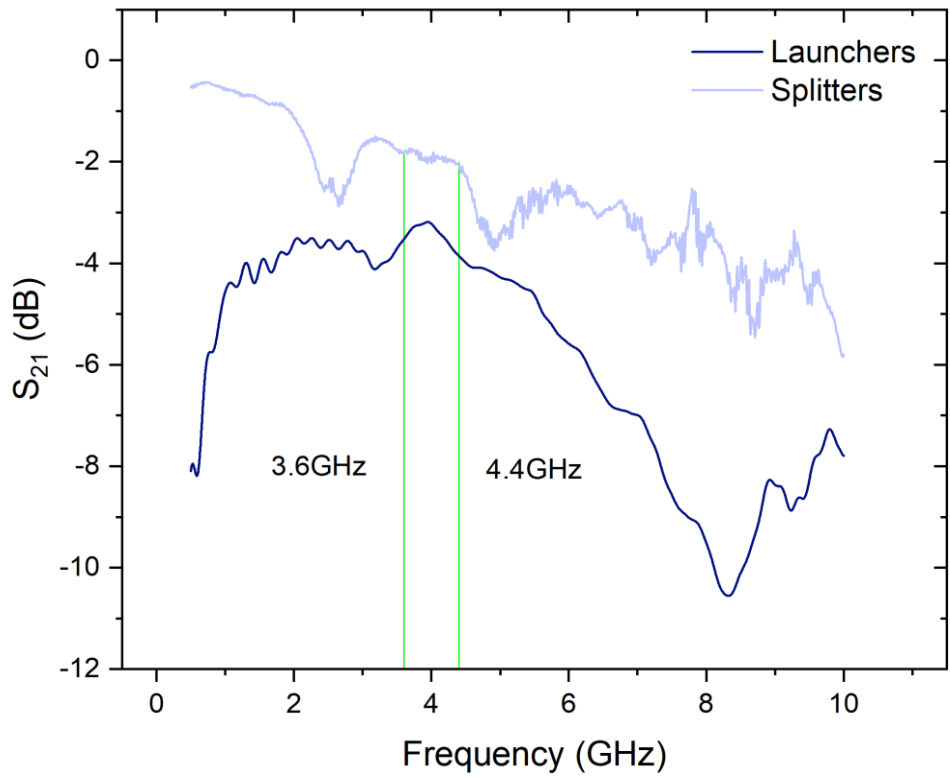

**Supplementary Figure 9** $S_{21}$ measurement results for the a pair of launchers and a pair splitters. The latter includes the losses of the 4 connecting rigid coaxial cables as well.